\documentclass[useAMS,usenatbib]{mn2e}

\newcommand\ionm[2]{#1$\;${\small\rmfamily{#2}}\relax}% 
\newcommand{\be}{\begin{enumerate}}
\newcommand{\ee}{\end{enumerate}}

\newcommand{\chisq}{\chi^2}

\newcommand{\tlogoh}{$12+\log(\mbox{O}/\mbox{H})$}
\newcommand{\msun}{\ifmmode {\rm M}_{\odot} \else M$_{\odot}$\fi}
\newcommand{\msfr}{\dot{M}_{\rm{SFR}}}
\newcommand{\mdotstar}{\dot{M}_{\star}}
\newcommand{\macc}{\dot{M}_{\rm acc}}
\newcommand{\macctot}{\dot{M}_{\rm acc,\,tot}}
\newcommand{\mwind}{\dot{M}_{\rm w}}
\newcommand{\mstar}{M_{\star}}
\newcommand{\mhalo}{M_{\rm halo}}
\newcommand{\etaa}{\eta_{\rm a}}
\newcommand{\etaw}{\eta_{\rm w}}
\newcommand{\zetaa}{\zeta_{\rm a}}
\newcommand{\zetaw}{\zeta_{\rm w}}
\newcommand{\zigm}{Z_{\rm IGM}}
\newcommand{\zej}{Z_{\rm ej}}
\newcommand{\zg}{Z_{\rm g}}
\newcommand{\zw}{Z_{\rm w}}
\newcommand{\mg}{M_{\rm g}}
\newcommand{\fe}{f_{\rm e}}
\newcommand{\fg}{F_{\rm g}}

\newcommand{\fcold}{f_{\rm cold}}
\newcommand{\frecy}{f_{\rm recy}}
\newcommand{\kms}{\ifmmode \mbox{\,km\,s}^{-1} \else \,km\,s$^{-1}$\fi}
\newcommand{\mug}{\mu_{\rm g}}
\newcommand{\vvir}{v_{\rm vir}}
\newcommand{\zmax}{Z_{\rm ej,max}}
\newcommand{\mecl}{M_{\rm ecl}}
\newcommand{\meclmax}{M_{\rm ecl,\,max}}
\newcommand{\mmax}{m_{\rm max}}

\newcommand{\mzr}{mass-metallicity relation}

\newcommand{\apj}{ApJ}
\newcommand{\apjl}{ApJL}
\newcommand{\apjs}{ApJS}
\newcommand{\aj}{AJ}

\newcommand{\mnras}{MNRAS}
\newcommand{\nat}{Nature}

\newcommand{\pasp}{PASP}
\newcommand{\araa}{ARA\&A}
\newcommand{\aap}{A\&A}

\newcommand{\physrep}{PhysRep}
\newcommand{\fcp}{FCPh}

\usepackage{graphicx}
\usepackage{amssymb}
\usepackage{verbatim}
\usepackage{times}
\usepackage[usenames]{color}

\begin{document}
\title[Constraining star formation driven galaxy winds]{Constraints on star formation driven galaxy winds from the
mass-metallicity relation at  $z=0$}

\author[Peeples \& Shankar]{Molly S.\ Peeples$^{1,2}$\thanks{E-mail:
    molly@astro.ucla.edu} \&\ Francesco
    Shankar$^{3}$\thanks{E-mail: shankar@mpa-garching.mpg.de} \\
$^{1}$Department of Astronomy and Center for Cosmology and
    Astro-Particle Physics, Ohio State University, 140 W.\ 18th Ave.,
    Columbus,~OH~43210\\
$^{2}$Southern California Center for Galaxy Evolution Fellow, University of
California Los Angeles, 430 Portola Plaza, Los~Angeles,~CA~90095\\
$^{3}$Max-Planck-Instit\"{u}t f\"{u}r Astrophysik,
    Karl-Schwarzschild-Str. 1, D-85748, Garching, Germany }

\pagerange{\pageref{firstpage}--\pageref{lastpage}} \pubyear{2010}

\date{\today}

\maketitle

\label{firstpage}

\begin{abstract}
  We extend a chemical evolution model relating galaxy stellar mass
  and gas-phase oxygen abundance (the \mzr) to explicitly consider the
  mass-dependence of galaxy gas fractions and outflows.  Using
  empirically derived scalings of galaxy mass with halo virial
  velocity in conjunction with the most recent observations of $z\sim
  0$ total galaxy cold gas fractions and the \mzr, we place stringent
  global constraints on the magnitude and scaling of the efficiency
  with which star forming galaxies expel metals.  We demonstrate that
  under the assumptions that metal accretion is negligible and the
  stellar initial mass function does not vary, efficient outflows are
  required to reproduce the \mzr; without winds, gas-to-stellar mass
  ratios $\gtrsim 0.3$\,dex higher than observed are needed.
  Moreover, $z=0$ gas fractions are low enough that while they have
  some effect on the magnitude of outflows required, the slope of the
  gas fraction--stellar mass relation does not strongly affect our
  conclusions on how the wind efficiencies must scale with galaxy
  mass. Because theoretical descriptions of the mass loading factor
  $\etaw\equiv\mwind/\msfr$, where $\mwind$ is the mass outflow rate
  and $\msfr$ is the star formation rate, are often cast in terms of
  the depth of the galaxy potential well, which is in turn linked to
  the host halo virial velocity $\vvir$, we use one of the latest
  abundance matching analyses to describe outflow efficiencies in
  terms of $\vvir$ rather than stellar mass.  Despite systematic
  uncertainties in the normalization and slope of the \mzr, we show
  that the metal expulsion efficiency $\zetaw\equiv(\zw/\zg)\etaw$
  (where $\zw$ is the wind metallicitiy and $\zg$ is the interstellar
  medium metallicity) must be both high and scale steeply with mass.
  Specifically, we show that $\zetaw\gg 1$ and
  $\zetaw\propto\vvir^{-3}$ or steeper.  In contrast, momentum- or
  energy-driven outflow models suggest that $\etaw$ should scale as
  $\vvir^{-1}$ or $\vvir^{-2}$, respectively, implying that the
  $\zw$-$\mstar$ relation should be shallower than the $\zg$-$\mstar$
  relation.
\end{abstract}


\section{Introduction}\label{sec:intro}
Star-forming galaxies follow a tight ($\sim 0.1$\,dex scatter)
correlation between their gas phase oxygen abundance (hereafter
referred to as ``metallicity'') and stellar mass \citep{tremonti04}.
This \mzr\ is primarily understood to be a sequence of oxygen {\em
  suppression}, rather than enrichment
\citep{tremonti04,dalcanton07,erb08,finlator08}.  The production of
oxygen traces the production of stars, implying that the observed
trend in the oxygen-to-gas ratio reflects either a trend in the galaxy
gas-to-stellar mass ratio or in processes that affect gas-phase metals
but not stars.  A consensus is emerging that although galaxy gas
fractions can and do affect the \mzr, if the stellar initial mass
function (IMF) is the same in all galaxies, then outflows that are
more efficient at removing metals from low-mass galaxies are required
in order to reproduce the observations
\citep[e.g.,][]{dalcanton07,finlator08,spitoni10}.  However, the
global properties of these outflows and the physics underlying how
star formation drives them are not well understood---and winds are
expensive and difficult to observe directly.  In this paper, we
incorporate the most recent observations of galaxy gas fractions and
the \mzr\ at $z\sim 0$ into a simple chemical evolution model to
explore what constraints can be placed on how the efficiencies and
composition of star formation driven galactic winds scale with galaxy
stellar mass and halo virial velocity.

Several analytic studies have concluded that star formation driven
outflows are crucial to reproducing the observed \mzr.  \citet{erb08}
used a simple analytic chemical evolution model to argue that the star
formation rate, $\msfr$, and the outflow rate, $\mwind$, should be
roughly equal. While $\mwind$ and the gas accretion rate $\macc$ vary
with the star formation rate (and thus gas fraction),
$\etaw\equiv\mwind/\msfr$ and $\etaa\equiv\macc/\msfr$ are constant
universal parameters, a common practice in analytic models of galaxy
chemical evolution \citep[see also][and references therein]{samui08}.
Though models specifically aimed at duplicating observations of the
\mzr\ commonly assume $\zw=\zg$, \citet{dalcanton07} argues that
metal-enriched outflows (those comprised predominantly of Type~II
supernova ejecta, and thus with $\zw>\zg$) are required if the rate of
gas accretion is to be reasonable.  More recently, \citet{spitoni10}
have argued that the $z=0$ \mzr\ together with gas fractions derived
by inverting the Kennicutt-Schmidt
\citep[K-S,][]{kennicutt98,schmidt59} law imply that not only are
outflows required, but that they must be more efficient at removing
metals from low-mass galaxies than from more massive ones.
\citet{finlator08} drew a similar conclusion by analyzing a suite of
cosmological smoothed particle hydrodynamics (SPH) simulations evolved
with {\sc Gadget-2} \citep{springel05} in conjunction with detailed
analytic models.  They showed that, in general, $\zg\propto\etaw^{-1}$
for $\etaw\gg 1$.  Their favored model that reproduces the
\citeauthor{erb06a}\ $z\sim 2.2$ \mzr\ is one in which $\etaw\propto
\sigma^{-1}$, where $\sigma$ is the galaxy velocity
dispersion.\footnote{This parameterization is motivated by the
  observations of \citet{martin05} and the theory of momentum-driven
  winds \citep{murray05}; see \S\ref{sec:winds} for more details.} In
this simulation, $\sigma^{-1}\propto M_{\rm
  halo}^{-1/3}\propto\mstar^{-1/3}$, which naturally explains why this
$\etaw$ scaling is able to reproduce a \mzr\ with
$\zg\propto\mstar^{0.3}$.  These simple scaling relations highlight a
link between a galaxy's stellar mass, its halo mass, and its potential
well: wind models aimed at successfully reproducing the \mzr\ also
need to correctly reproduce (or incorporate) the $\mstar$-$\mhalo$\
relation.  Moreover, this analysis shows that the \mzr\ is a
potentially powerful tool for constraining how star formation driven
outflows scale with galaxy and halo properties; this is particularly
interesting as such scalings are currently not well constrained
through either direct observations or theoretical considerstions.

On the other hand, several models focus instead on the efficiency of
star formation as a function of stellar mass.  In such models, an
increase in the star formation efficiency with galaxy mass---without the
need for outflows---is sufficient to reproduce the observed \mzr\
\citep{calura09a}.  \citet{brooks07} used a set of SPH simulations
evolved with Gasoline \citep{wadsley04}---and therefore a different
recipe for star-formation feedback\footnote{Because of the resolution of
cosmological SPH simulations, star-formation feedback must be included
using ``recipes'' instead of directly modelling the underlying physics.
The winds in \citeauthor{finlator08}'s simulations are implemented by
physically moving gas particles away from star-forming regions.  In
\citeauthor{brooks07}'s simulations, star formation thermally heats
neighboring particles.  In both prescriptions, the relevant particles
are not allowed to interact hydrodynamically (\citeauthor{finlator08})
or radiatively cool (\citeauthor{brooks07}) for some
physically-motivated amount of time.} than \citet{finlator08}--- to
argue that preferentially expelling gas from the low-mass galaxies is
insufficient for reproducing the observed \mzr.  These authors claim
that it is instead the reduced star-formation efficiency (and thus
differences in galaxy gas fractions) induced by such feedback that is
primarily responsible for driving the relation's morphology.  In the
context of the \mzr, variations in the star formation efficiency affect
galaxy gas fractions (as well as the $\mstar$-$\mhalo$\ relation).  
We do not directly address star formation efficiency here because we
take both galaxy gas fractions and the $\mstar$-$\mhalo$ relation as
givens rather than something to be constrained by the model; we
discuss in Appendix~\ref{app:gas} the implications our choices for
these relations have on how star formation efficiency varies with
galaxy mass.

Finally, letting the IMF (and thus the amount of oxygen produced per
unit stellar mass) vary with galaxy mass provides a straightforward
way to reproduce the \mzr\
\citep{tinsley74,koppen07,recchi09,calura09b,spitoni10} .  We
primarily assume here that the IMF is the same in all galaxies, with a
brief discussion in \S\,\ref{sec:igimf} of how uncertainties in yields
in the presence of a variable IMF affect our results.  In general, if
the IMF is top-light in low-mass galaxies then this will imply that
outflow efficiencies do not need to scale as steeply with mass as
suggested by the non-varying IMF case.

We apply here a simple model with which to understand the \mzr\ to the
\mzr\ at $z\sim 0$, where external constraints such as gas fractions
and the stellar mass function are best measured.  We do not assume a
particular form for the \mzr; we instead base our conclusions on the
range of parameter space allowed by the range of systematic
uncertainties in interpretting strong nebular emission lines as oxygen
abundances (\S\,\ref{sec:mzrs}).  Our main simplifying assumptions are
that the metallicity of gas accreted from the intergalactic medium is
negligible and that the nucleosynthetic yield is constant with galaxy
mass.  With these constraints and assumptions, the only free
parameters are those describing outflows, which we are able to
describe as a function of halo virial velocity.

This paper is organized as follows.  In \S\,\ref{sec:obs}, we discuss
the relevant observations.  The slope and normalization of the \mzr\
strongly affect the interpretted properties of galaxy winds.
Unfortunately, while there exist exquisite data on emission line
ratios of star forming galaxies at $z=0$, the correct way to interpret
these line ratios in terms of oxygen abundances is not agreed upon; we
therefore consider several measurements of the $z=0$ \mzr, as outlined
in \S\,\ref{sec:mzrs}.  It is commonly assumed in chemical evolution
models, and we assume here, that the gas is well-mixed; we address the
differences between galaxies' cold gas resevoirs and the gas traced by
star formation in \S\,\ref{sec:gas} (see also Appendix~\ref{app:ks}).
As the purpose of this paper is to place constraints on how galaxy
outflows scale with galaxy mass, we briefly outline observed
properties of galaxy outflows (and theoretical models thereof) in
\S\,\ref{sec:winds}.  We lay out the formalism in \S\ref{sec:modmzr}
and its derivation in Appendix~\ref{app:mzr}, along with how we
connect galaxy stellar masses to host halo properties
(\S\,\ref{sec:halos}).  In \S\,\ref{sec:results}, we show how gas
dilution and outflows must combine in order to yield the observed
\mzr, and what this implies about galaxy outflows in order for
predicted gas fractions to be consistent with the data; further
details are presented in Appendix~\ref{app:results}. How these
conclusions are affected by uncertainties in the yield is addressed in
\S\,\ref{sec:igimf}.  We then present in \S\,\ref{sec:zwind} what
constraints wind metallicity and entrainment fraction considerations
place on viable outflow models, with a summary and further discussion
in \S\,\ref{sec:disc}.  Appendix~\ref{app:gas} describes the
connection between gas masses, accretion, and star formation rates in
our approach, with implications for star formation efficiency.

Throughout we adopt a cosmology of
$(\Omega_m,\Omega_b,\sigma_8,h)=(0.26,0.047,0.77,0.72)$ and a
\citet{chabrier03b} initial mass function (IMF), unless otherwise noted.
Varying the cosmological parameters within the ranges allowed from
observations \citep[e.g.,][]{hinshaw09} does not alter our conclusions.
The impact of varying $\Omega_m$ or $\Omega_b$, has, for example, little
effect on the shape of the $\mstar$-$\mhalo$ relation or on the
determination of the stellar masses in SDSS. Though varying $\sigma_8$
does change the number density of massive halos, it has little impact on
the range of halo masses of interest here. Finally, we note that the
virial relations only have a mild change in normalization when varying
cosmological parameters, without having much impact on our overall
results.

\section{Relevant Observations}\label{sec:obs}
\subsection{The observed $z\sim 0$ mass-metallicity relation}\label{sec:mzrs}
Since oxygen is effectively produced only in Type~II SNe---the deaths
of massive, short-lived stars---and \ionm{H}{II}\ regions are
associated with ongoing star formation, the gas-phase
``mass-metallicity relation'' typically refers to only the galaxy's
oxygen abundance in gas that is currently forming stars; we therefore
will use ``metals'' and ``oxygen'' interchangeably unless otherwise
noted.  However, though \tlogoh\ is measured at the sites of star
formation, the measured abundances are the {\em birth} abundances of
the \ionm{H}{II}\ regions; supernovae (the sites of oxygen production)
destroy their nascent clouds, rendering so-called ``self enrichment''
of \ionm{H}{II}\ regions extremely rare.  We therefore assume that the
galaxy gas is well-mixed, i.e., that the mixing time is short relative
to the timescale for star formation.

\begin{figure}
\includegraphics[width=0.48\textwidth]{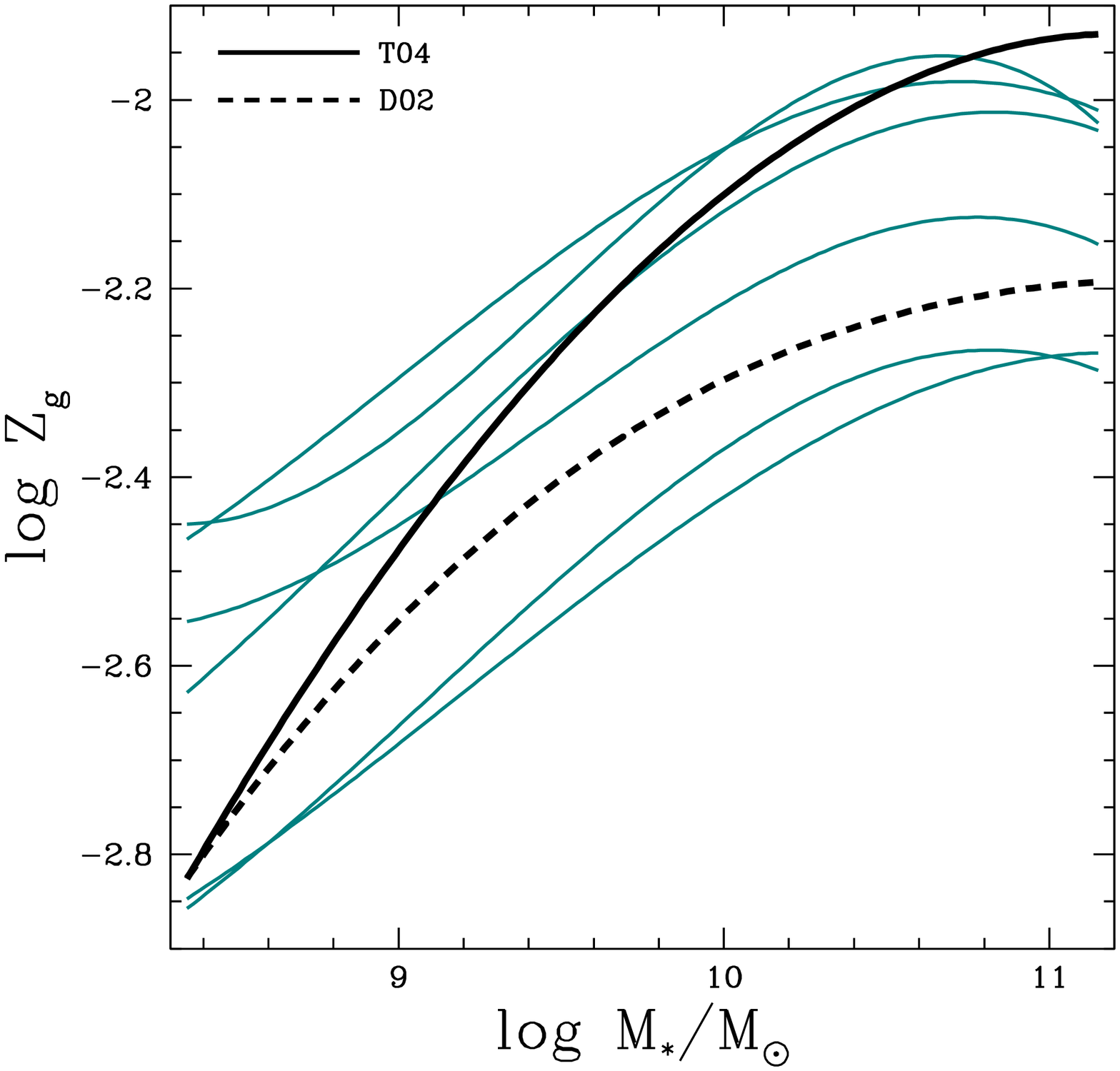}
\caption{\label{fig:ke08}Mass-metallicity relations listed in
  Table~\ref{tbl:ke08mz} \citep[][and
  equation~\ref{eqn:lzg}]{kewley08}.  The scatter about any given one
  of these curves is $0.1$--$0.15$\,dex, which is much less than the
  differences in normalization; that is, the normalization differences
  are systematic.  The mass-metallicity relations in black (T04,
  solid; D02, dashed) are modeled in detail \S\S\,\ref{sec:results},
  \ref{sec:zwind}, and Appendix~\ref{app:results}.  }
\end{figure}

\begin{table}
\centering
\scriptsize
\begin{tabular}{lrrrr}\hline\hline 
ID & $a$ & $b$ & $c$ & $d$\\\hline\hline 
{\bf T04} & $\mathbf{-0.759210}$ & $\mathbf{1.30177}$ & $\mathbf{0.003261}$ & $\mathbf{-0.00364112}$ \\ 
Z94 & $73.0539$ & $-20.9053$ & $2.23299$ & $-0.0783089$ \\ 
KK04 & $28.1404$ & $-7.02595$ & $0.812620$ & $-0.0301508$ \\ 
KD02 & $28.4613$ & $-7.32158$ & $0.855119$ & $-0.0318315$ \\ 
M91 & $46.1480$ & $-12.3801$ & $1.33589$ & $-0.0471074$ \\ 
{\bf D02} & $\mathbf{-8.91951}$ & $\mathbf{4.18231}$ & $\mathbf{-0.323383}$ & $\mathbf{0.00818179}$ \\ 
PP04O3N2 & $32.5769$ & $-8.61049$ & $0.981780$ & $-0.0359763$ \\ 
PP04N2 & $24.1879$ & $-5.69253$ & $0.648668$ & $-0.0235065$ \\ 
\end{tabular}
\caption{\citet{kewley08} fits to the mass-metallicity relation, where
  $\log \zg = a + b\log\mstar + c(\log\mstar)^2 + d(\log\mstar)^3$,
  sorted by decreasing $\max(\zg)$.  The two fits we consider in the
  main text (T04 and D02) are in bold.  See text for abbreviations.
\label{tbl:ke08mz}}
\end{table}

Observationally, oxygen abundance increases with galaxy stellar mass.
This relation has very little scatter ($\sim 0.1$\,dex in
$12+\log[\mbox{O}/\mbox{H}]$ at fixed stellar mass), though severe
outliers do exist \citep{peeples08,peeples09a}.  The amplitude and
slope of the \mzr, however, are not well constrained, despite
exquisite and extensive data from the Sloan Digital Sky Survey
\citep[SDSS; ][]{adelman06}.  This ambiguity is due to the theoretical
uncertanties in how to convert emission-line fluxes to \tlogoh, as
assumptions must be made about both the gas temperature and ionization
structure.  While the electron temperature can be estimated directly
using the [\ionm{O}{III}]\,$\lambda 4363$ auroral line, this line is
extremely weak and usually only detectable in very metal-poor
environments.  Thus, it is common to calibrate measurement methods
using much stronger forbidden emission lines such as
[\ionm{O}{II}]\,$\lambda\lambda 3726,3729$, H$\beta$,
[\ionm{O}{III}]\,$\lambda\lambda 4959,5007$, H$\alpha$, and
[\ionm{N}{II}]\,$\lambda 6584$ based on the so-called direct
[\ionm{O}{III}]\,$\lambda 4363$ $T_e$ method.  However, since
[\ionm{O}{III}]\,$\lambda 4363$ preferentially emits in
high-temperature regions, this calibration can lead to an
over-estimate of the electron temperature based on this line and thus
an under-estimate of the oxygen abundance \citep{kewley08}.  It is
therefore common to instead calibrate strong-line measurement methods
based on theoretical photoionization models.  On the other hand, there
are arguments that such strong-line methods {\em over}-estimate the
true abundance \citep{kennicutt03}.  Moreover, most indicators are
either double-valued at low metallicities (such as the popular
$R_{23}$ indicator) or saturate at high metallicites as emission-line
cooling shifts to the near-infrared \citep{bresolin06}.

\citet{kewley08} highlight many of these issues, and derive \tlogoh\ for
a large set of galaxies from SDSS using ten indicators (eight of which
we consider here: T04, \citealt{tremonti04}; D02, \citealt{denicolo02};
KK04, \citealt{kobulnicky04}; Z94, \citealt{zaritsky94}; KD02,
\citealt{kewley02}; M91, \citealt{mcgaugh91}; PP04O3N2 and PP04N2, using
the \citealt{pettini04}
([\ionm{O}{III}]/H$\beta$)/([\ionm{N}{II}]/H$\alpha$) and
[\ionm{N}{II}]/H$\alpha$ flux ratios, respectively).  The
\citeauthor{kewley08}\ fits to the \mzr\ are given in
Table~\ref{tbl:ke08mz}, where we have converted from a \citet{kroupa01}
to a \citet{chabrier03b}\ IMF and from \tlogoh\ to $\log \zg$, where
\begin{eqnarray}\label{eqn:lzg}
\log \zg &=& [12+\log(\mbox{O}/\mbox{H})] - 12 - \log\left[\frac{M_{\rm O}/M_{\rm H}}{XM_{\rm H}+YM_{\rm He}}\right]\\
&=& \log(\mbox{O}/\mbox{H})- \log\left[\frac{15.999/1.0079}{0.75\times1.0079+0.25\times4.0026}\right].\nonumber
\end{eqnarray}
These mass-metallicity relations are plotted in Figure~\ref{fig:ke08};
the scatter in $\zg$ at fixed $\mstar$ for each mass-metallicity
relation is smaller by a factor of 2--3 than the spread in
normalizations, implying that the differences are caused by the
systematics discussed above.  

We consider in detail the two relations in black in
Figure~\ref{fig:ke08} and in bold in Table~\ref{tbl:ke08mz} (T04,
\citealt{tremonti04} and D02, \citealt{denicolo02}).  The D02
indicator is a linear relation between the [\ionm{N}{II}]$\,\lambda
6584$/H$\alpha$ ratio and \tlogoh calibrated against $T_e$
metallicities.  The relatively low normalization of this method is
common for $T_e$-calibrated indicators.  The T04 method is based on
theoretical stellar population synthesis and photoionization models
combined with a Bayesian analysis of many more strong emission lines
than used in most methods.  While we do not favor any one \tlogoh\
indicator, we take these two mass-metallicity relations as
representative of the normalizations and slopes observations as a
whole.

\subsection{Observed gas fractions of $z\sim 0$ galaxies}\label{sec:gas}
\begin{figure*}
\includegraphics[width=0.495\textwidth]{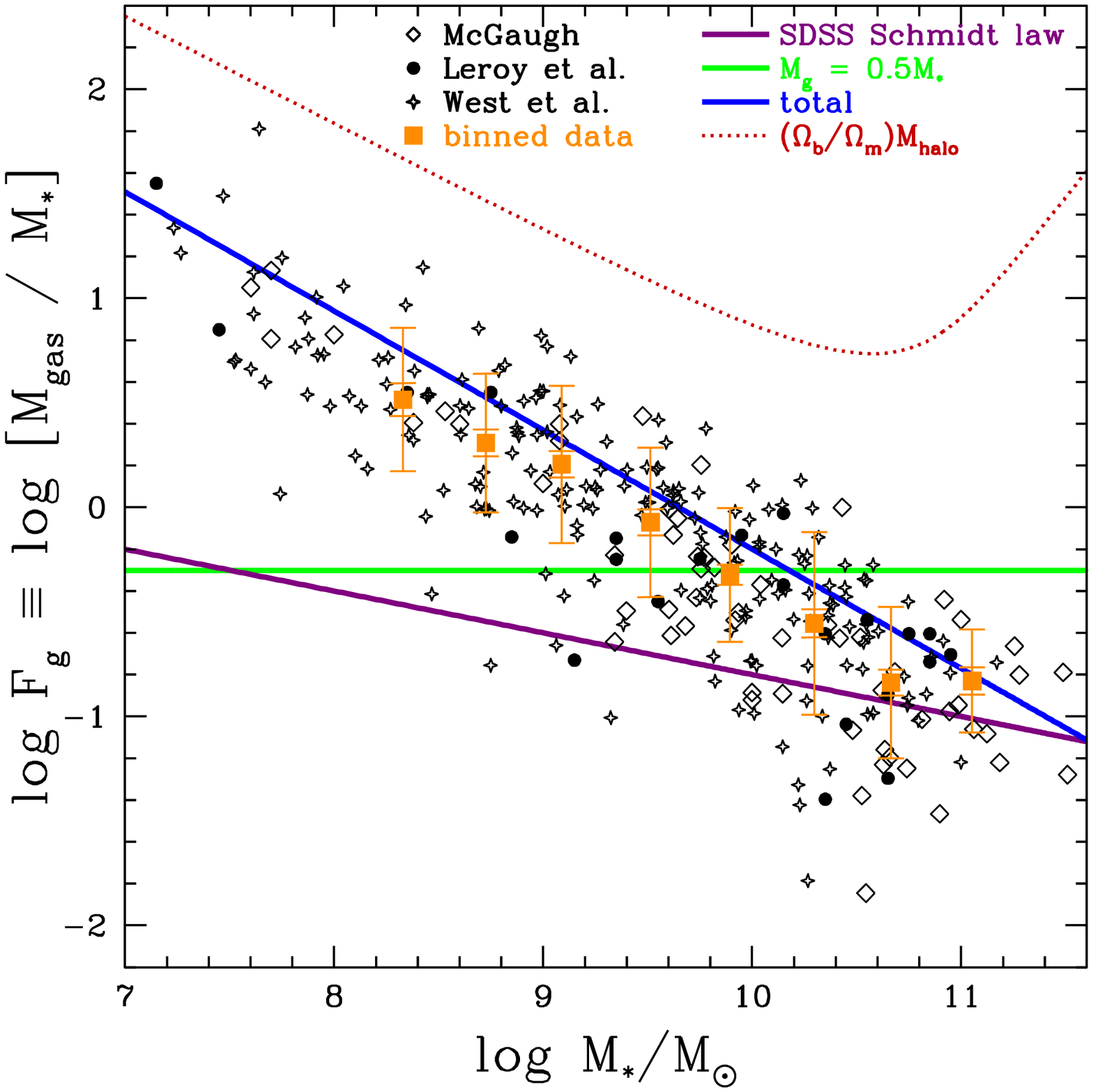}
\hfill
\includegraphics[width=0.495\textwidth]{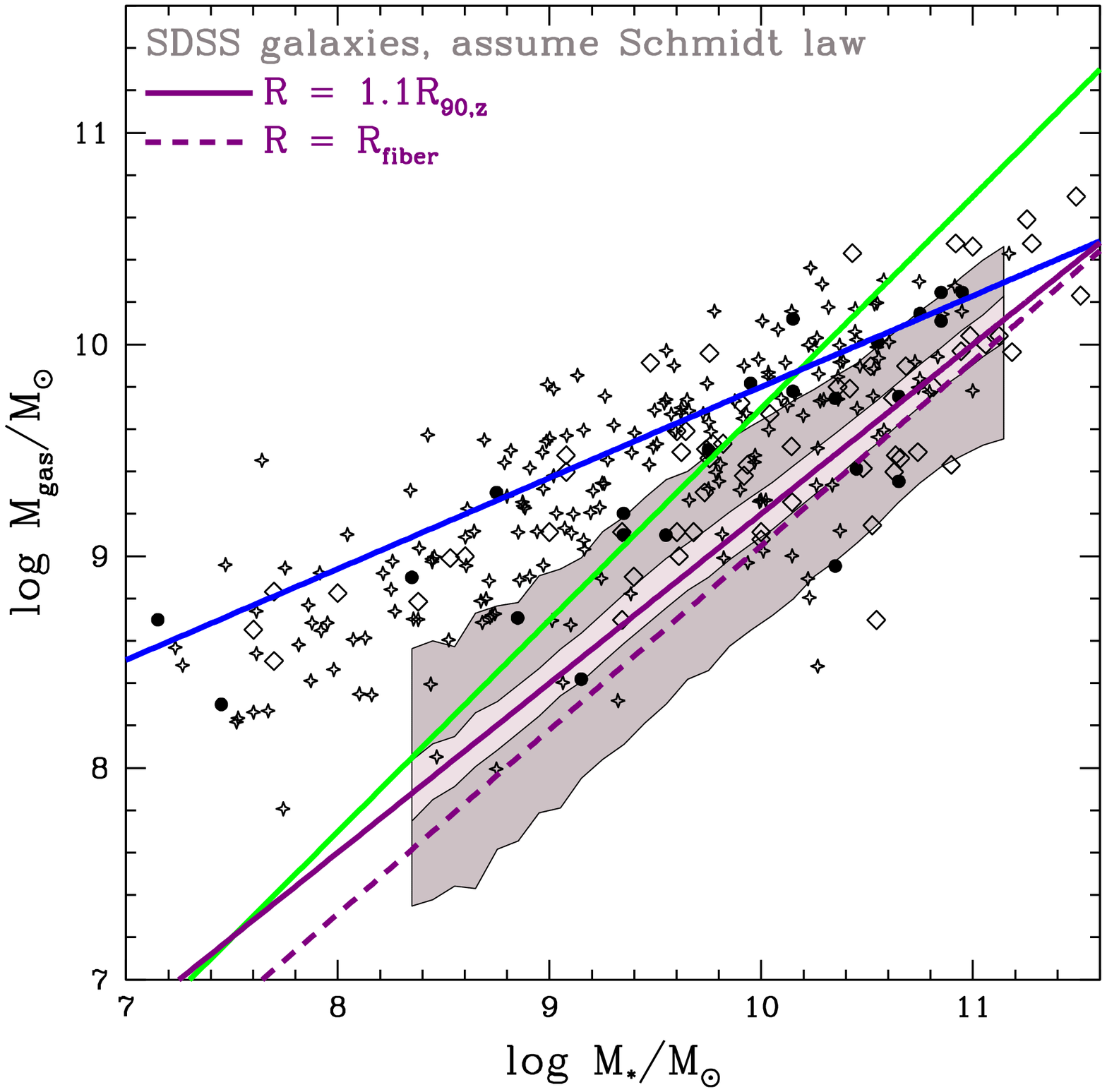}
\caption{\label{fig:Fg}Left: Gas fractions $\fg$ as a function of
  $\mstar$. Right: Gas masses $\mg$ as a function of $\mstar$.  The open
  black diamonds are \ionm{H}{I}\ gas fractions and masses from
  \citet{mcgaugh05}; the crosses are the same from
  \citet{west09,west10}.  The filled circles are \ionm{H}{I}~+~H$_2$ gas
  fractions and masses \citep{leroy08}, who find that there is very
  little H$_2$ below $\log\mstar\sim 9.5$, which is consistent with the
  comparison to the \ionm{H}{I}\ samples.  The red dotted line shows the
  maximum baryonic mass $(\Omega_b/\Omega_m)\mhalo$, while the green
  ``flat'' line shows $\mg=0.5\mstar$.  The blue ``total'' line is a fit
  to these data with the normalization increased by 0.2\,dex; the orange
  squares are the mean $\log\fg$ of these same data in bins of
  $\Delta\log\mstar=0.4$\,dex with the inner and outer errorbars
  denoting the uncertainty in and dispersion about the mean,
  respectively. Gas fractions and masses derived from SDSS data and
  inverting the K-S law, assuming a radius of $1.1R_{90,z}$ (solid line)
  and the fiber radius (dashed line); in the right panel, the shaded
  region corresponds to the 1- and 2-$\sigma$ dispersions in moving bins
  of $\log\mstar$.  }
\end{figure*}

Figure~\ref{fig:Fg} shows how the gas-to-stellar mass ratio $\fg$
(left panel) and gas mass $\mg$ (right panel) vary with galaxy stellar
mass.  The open diamonds are total \ionm{H}{I}\ gas masses measured
from 21\,cm line fluxes \citep{mcgaugh05}. The crosses are also
\ionm{H}{I}\ gas masses, with stellar masses measured from SDSS
\citep{garcia09,west09,west10}.  The filled circles represent the
total \ionm{H}{I}~+~H$_2$ gas masses (including a correction for
helium) from The \ionm{H}{I}\ Nearby Galaxy Survey (THINGS), with the
H$_2$ masses derived from HERA CO-Line Extragalactic Survey (HERACLES)
and the Berkeley-Illinois-Maryland Association Survey of Nearby
Galaxies (BIMA SONG) CO measurements \citep{leroy08}.  Though there is
large scatter in the gas fraction at a fixed stellar mass, gas
fractions clearly decrease as $\mstar$ increases; this behavior is
found in cosomological hydrodynamic simulations
\citep[e.g.,][]{hopkins08}.  The mean $\log\fg$ in bins of
$\Delta\log\mstar=0.4$\,dex for $8.1\leq\log\mstar\leq 11.3$ is
overplotted with the large solid orange squares; we list these means
and uncertainties in Table~\ref{tbl:Fgbins}.  Each of these data sets
focus on star-forming galaxies similar to those in which \tlogoh\ is
measurable; surveys not restricted to actively star-forming galaxies
lead to much lower average gas fractions \citep{catinella10}.

\begin{table}
\centering
\begin{tabular}{rrr}\hline\hline 
$\langle{\log\mstar}\rangle$ & $\langle{\log\fg}\rangle$ & $\sigma_{\log\fg}$\\\hline
$8.3298$ & $0.5153$ & $0.07867$  \\ 
$8.7265$ & $0.3084$ & $0.06500$  \\ 
$9.0892$ & $0.2062$ & $0.06359$  \\ 
$9.5141$ & $-0.07142$ & $0.06220$  \\ 
$9.8941$ & $-0.3230$ & $0.04817$  \\ 
$10.298$ & $-0.5548$ & $0.06666$  \\ 
$10.664$ & $-0.8389$ & $0.06212$  \\ 
$11.053$ & $-0.8303$ & $0.06566$  \\ \hline 
\end{tabular}
\caption{Cold gas fractions $\log\fg = \log(\mg/\mstar)$ in bins of
$\Delta\log\mstar=0.4$\,dex and the uncertainty in the mean
$\sigma_{\log\fg}$ for the \citet{mcgaugh05}, \citet{leroy08}, and
\citet{garcia09} data sets.
\label{tbl:Fgbins}}
\end{table}

We parameterize $\fg$ as power-law of the form
\begin{equation}\label{eqn:Fg}
\fg \equiv \frac{\mg}{\mstar} = \left(\frac{\mstar}{M_{\star,0}}\right)^{-\gamma}
= K_f\mstar^{-\gamma},
\end{equation}
with $\gamma > 0$. Table~\ref{tbl:Fg} lists $\log M_{\star,0}$, $K_f$
and $\gamma$ for our adopted gas relations.  As we show in
\S\ref{sec:modmzr}, $\fg$ is a more convenient parameterization than the
commonly used and more arguably intuitive $\mug$, the gas mass as a fraction of
the total baryonic galaxy mass $\mstar+\mg$,
\begin{equation}
\mug\equiv \frac{\mg}{\mg+\mstar}=\frac{\fg}{1+\fg}.
\end{equation}
The ``total'' gas fraction relation is a power-law fit to the combined
\citeauthor{mcgaugh05}, \citeauthor{leroy08}, and \citeauthor{garcia09}\
data sets, offset by $+0.2$\,dex so that the total gas fractions are
greater than those implied by the K-S law (see below).  In order to
understand the contribution of a sloped gas fraction relation to the
\mzr, we also consider a flat gas relation of $\mg=0.5\mstar$, shown in
green in Figure~\ref{fig:Fg}.

\begin{table}
\centering
\begin{tabular}{lrlr}\hline\hline 
Name & $\log M_{\star,0}$ & $K_f$ & $\gamma$\\\hline
Total & $9.6$ & $316228$ & $0.57$ \\
SDSS & $6.0$ & $15.85$ & $0.20$ \\
Fiber & $2.7$ & $2.24$ & $0.13$ \\
Flat & --- & $0.50$ & $0.00$ \\\hline
\end{tabular}
\caption{Gas fraction relation parameters, $\fg = \mg/\mstar = K_f \mstar^{-\gamma}$.
\label{tbl:Fg}}
\end{table}

For reference, Figure~\ref{fig:Fg} shows how the total baryonic halo
mass, $(\Omega_b/\Omega_m)M_{\rm h}$, varies with stellar mass (halo
mass as a function of $\mstar$ is calculated as discussed in
\S\ref{sec:halos}).  The offset between the baryonic halo mass and
$\mstar+\mg$ is evidence of the so-called ``missing baryon'' problem;
the missing baryons are either hot or have been expelled from the
halos by $z=0$ \citep{crain07}.  Figure~\ref{fig:Fg} further
highlights the fact that for $\mstar\lesssim 10^{10}\,\msun$, the
fraction of baryons in the form of cold gas is roughly constant (i.e.,
the blue and red lines are roughly parallel).  Moreover, while massive
galaxies are gas poor, galaxies with stellar masses below $\sim
10^{9.5}\,\msun$ have most of their mass in the form of gas: the
processes responsible for the missing baryons in $z=0$ halos must also
account for this inefficiency of star formation in low mass halos.  We
discuss this issue further in Appendix~\ref{app:gas}.

The solid line in the right panel of Figure~\ref{fig:Fg} shows the
median gas fractions obtained by inverting the Kennicutt-Schmidt
\citep[K-S,][]{kennicutt98,schmidt59} relation, as explained in detail in
Appendix~\ref{app:ks}.  The shaded contours denote the 1-$\sigma$ and
2-$\sigma$ gas masses derived for the entire galaxy ($R_g = 1.1R_{90,
  z}$) in running bins of $\log\mstar$ from $\log\mstar=8.3$ to 11.1;
for clarity, galaxies falling outside this region are not shown.  The
solid line is an eyeball power-law fit to the median $R_g = 1.1R_{90,
  z}$ gas masses while the dashed line is the same for the gas (and
stellar) masses within the SDSS fiber. The fact that these relations
are quite similar to one another indicates that aperture corrections
are relatively small and/or that gas fractions are relatively
scale-invariant within $1.1R_{90, z}$.

The gas masses estimated from the K-S law and the measurements of
total cold gas masses roughly agree with one another on the low gas
fraction of $\fg\sim 0.1$ at $\log\mstar\sim 11$, and that $\fg$
increases with decreasing stellar mass.  The amount of this increase in
gas fraction, however, is in stark disagreement, with a range of over an
order of magnitude in $\fg$.  The K-S law only traces star-forming
gas and therefore traces molecular gas more closely than atomic, and
dwarf galaxies are deficient in molecular gas \citep{leroy08}.  At large radii
in more massive galaxies, the gas is predominately atomic, i.e., the
\ionm{H}{I}\ radii of galaxies is often much larger than the optical
(star-forming) radii \citep{boomsma08,walter08}.  For the purposes of
the \mzr, what matters is the total amount of gas that is able to
effectively mix and dilute metals.  A lower limit to this gas mass is
the gas that is able to collapse and form stars---the gas traced by the
K-S law.  If on the other hand the atomic and molecular gas are well
mixed (as opposed to, e.g., molecular gas only populating the galaxy
center and atomic gas being at large radii), then the total gas
fractions are more applicable.  Finally, neither of these gas fraction
estimates include ionized gas; if such gas is not only prevalent in
typical galaxies but also has efficient mass transfer with both supernova
ejecta and gas that will cool to form molecular clouds (and subsequently
\ionm{H}{II}\ regions), then even the ``total'' gas fraction relation
will be an underestimate of the gas diluting the galaxies' metals.

\subsection{Galaxy outflows}\label{sec:winds}
Though observations of galaxy-scale outflows are notoriously
difficult, galaxy winds observed in a range of star-forming galaxies
display a complex, multiphase structure.  Since detectability
increases with the star formation rate density \citep{veilleux05},
however, the most detailed studies of galaxy winds have been of the
outflows associated with extreme starbursts, namely, (ultra)luminous
infrared galaxies ([U]LIRGs).  Studies of blue-shifted
absorption-lines reveal both neutral
\citep{heckman00,rupke02,martin05} and photoionized gas
\citep{grimes09}, often with several kinematically distinct
components.  In contrast, X-ray emission around local starbursts such
as M82 indicates a hot ($T\sim 10^{6.5}$--$10^8$\,K), tenuous ($n\sim
10^{-4}$--$10^{-3}$\,cm$^{-3}$) wind fluid
\citep{strickland00,strickland07,strickland09}.  Wind velocities
derived from both emission and absorption line studies are typically
hundreds of km\,s$^{-1}$ \citep{martin05,grimes09}.  The outflow
velocity $v_{\rm w}$ of the colder neutral gas is typically comparable
to one to a few times the galaxy's circular velocity $v_{\rm circ}$
\citep{martin05}, which is comparable to the galaxy's virial velocity
$\vvir$ \citep[e.g.,][]{diemand07}.

The scaling $v_{\rm w}\sim\vvir$ follows naturally if momentum transfer from
radiation pressure is driving the wind \citep{murray05}.  For radiation
pressure to be effective, the starburst must be Eddington limited and
the outflowing gas has an asymptotic velocity of
\begin{equation}
v_{\rm w}(\infty)=2v_{\rm esc}\left(\frac{L}{L_{\rm edd}}-1\right)^{1/2},
\end{equation}
where the escape velocity $v_{\rm esc}$ is comparable to the virial
velocity.  The wind velocity is therefore typically taken to be
$v_{\rm w}=3\vvir$.  In the single-scattering limit \citep{murray05},
\begin{equation}\label{eqn:mom}
\mwind v_{\rm w} = \frac{L_{\rm starburst}}{c} = \frac{\epsilon_{\rm nuc}\msfr c^2}{c},
\end{equation}
where $L_{\rm starburst}$ is the starburst luminosity and $\epsilon_{\rm
  nuc}=8\times10^{-4}$ is the nuclear burning efficiency.  Thus the
  mass-loading factor\footnote{Definitions in the literature of the
  ``mass-loading factor'' vary; we take it to mean the {\em total}
  outflow mass rate divided by the {\em total} star formation rate
  (including short-lived stars).} $\etaw$ is proportional to the inverse
  of the virial velocity such that
\begin{equation}\label{eqn:etawmom}
\etaw\Big|_{\mbox{momentum}}\equiv\frac{\mwind}{\msfr}=\frac{\epsilon_{\rm nuc}
  c}{v_{\rm w}}\sim\frac{80\,{\rm km}\,{\rm s}^{-1}}{\vvir}.
\end{equation}
This same scaling is achieved if the wind is driven by cosmic rays
\citep{socrates08}.

On the other hand, the outflow may be driven by energy transfer, perhaps
from supernovae thermally heating the ISM \citep{chevalier85,
dekel86,silk98,murray05}.  In this popular scenario,
\begin{eqnarray}\label{eqn:energy}
\frac{1}{2}\mwind v_{\rm w}^2&\approx&\xi E_{\rm SN}\times\\ 
&&[\mbox{\# of SNe per solar mass of stars formed}]\msfr, \nonumber
\end{eqnarray}
where $E_{\rm SN}\sim 10^{51}$\,erg is the typical energy per supernova
and $\xi$ is the efficiency with which supernovae transfer energy to the
ISM.  Letting $\xi=0.1$, i.e., a 10\% efficiency, and taking the number
of supernovae per unit mass to be $10^{-2}$, this yields a mass-loading
factor of
\begin{equation}\label{eqn:etawenergy}
\etaw\Big|_{\mbox{energy}}\equiv\frac{\mwind}{\msfr}\sim\left(\frac{73\,{\rm km}\,{\rm s}^{-1}}{\vvir}\right)^2,
\end{equation}
where we have implicity assumed $v_{\rm w}\approx 3\vvir$.  While we in
general consider models in which $\etaw\propto\vvir^{-\beta}$ for
$\beta>0$ (or, equivalently, $\etaw\propto \mhalo^{-\beta/3}$, see
\S\ref{sec:halos}); it is helpful to keep the normalizations suggested
by equations~(\ref{eqn:etawmom}) and (\ref{eqn:etawenergy}) in mind.

Except via the impact of outflows on galaxy gas fractions (see
Appendix~\ref{app:gas}), the \mzr\ is insensitive to the {\em total}
mass outflow rate $\mwind$.  Instead, as we show in
\S\ref{sec:modmzr}, oxygen depletion due to winds is governed by the
rate of metal loss, $\zw\mwind$, where $\zw$ is the metallicity of the
outflow; in our case (see \S\ref{sec:mzrs}), the mass ratio of oxygen
in the outflowing material.  While many metals (oxygen, as well as,
e.g., iron, sodium, carbon, magnesium, and neon;
\citealt{heckman00,martin05,strickland07,martin09,grimes09,spoon09})
are observed in galaxy outflows, there are relatively few observations
of outflowing oxygen, and elemental abundances in the wind fluid are
rarely reported.  \citet{strickland09}, however, find that the X-ray
emitting outflow from M82 has a high enough metal content that it is
consistent with containing nearly all of the freshly produced metals
in the starburst with an inferred velocity of $\sim
1000$--2000\kms. Combined with their interpretation that the outflow
has very little entrained gas (i.e., that it is essentially comprised
solely of supernova ejecta), this implies that the metallicity of the
outflow is quite high.  (We note that in this interpretation of the
data, supernova explosions surprisingly have no radiative energy
losses when interacting with the ambient ISM [$\xi=1$ in
equation~\ref{eqn:energy}]; see also \citealt{heckman03}.)  This
picture is further complicated by the fact that outflows are likely
multi-phase, and the metallicities and escape fractions in, e.g., the
cold and ionized phases may be different. From the perspective of the
\mzr, however, what matters is the total amount of expelled oxygen
relative to the total amount of expelled gas, where ``expelled''
oxygen or gas is just the oxygen or gas that has either been
physically ejected from the galaxy or simply heated up such that it
cannot efficiently transfer mass to the gas that is able to cool and
form stars and thus be observed contributing to the \mzr.

\section{The formalism}\label{sec:model}
\subsection{The \mzr}\label{sec:modmzr}
The three galaxy masses relevant to the \mzr\ are the total galaxy
mass in stars, $\mstar$, the galaxy gas mass, $\mg$, and the mass of
gas-phase metals, $M_Z$.  The model is based on relating the
instantaneous change in these masses via their sources and sinks to
the instantaneous galaxy star formation rate, $\msfr$, ignoring
environmental effects such as mergers and tidal stripping \citep[see
also,
e.g.,][]{tinsley74,tinsley80,matteucci02,recchi09,finlator08,spitoni10}.
Observationally, \citet{mannucci10} and \citet{laralopez10} have
recently shown that $\zg$ has less scatter at fixed $\mstar$\ and
$\msfr$\ than at just fixed $\mstar$ (i.e., the \mzr); there is no
evidence for evolution of this surface up to $z\sim 2.5$.  This
finding implies that the $\mstar$-$\msfr$-$\zg$ hypersurface provides
a more physical description of the underlying physics than just the
$\mstar$-$\zg$ plane. In the formalism, the star formation rate is
closely linked with outflow efficiencies, and observationally, gas
fractions and star formation rates are tightly correlated.  We review
the relevant equations here and their derivation in
Appendix~\ref{app:mzr}.

The \mzr\ is described as
\begin{eqnarray}
\zg &=& y\Big[\zetaw-\zetaa + \label{eqn:zgfull}\\
    &&\quad\fg(1-\frecy)\left(\frac{{\rm d}\log
      \mg}{{\rm d}\log\mstar} + \frac{{\rm d}\log \zg}{{\rm
	d}\log\mstar}\right) + 1\Big]^{-1}\nonumber\\
&&\nonumber\\
 &=& y\left[\zetaw - \zetaa + \alpha \fg + 1\right]^{-1}, \label{eqn:zgsimple}
\end{eqnarray}
where 
\begin{equation}\label{eqn:alpha}
\alpha\equiv(1-\frecy)\left(\frac{{\rm d}\log
      \mg}{{\rm d}\log\mstar} + \frac{{\rm d}\log \zg}{{\rm
	d}\log\mstar}\right)
\end{equation}
is a factor of order unity.  Equation~(\ref{eqn:zgsimple}) shows that
the metallicity $\zg$\ is proportional to the nucleosynthetic yield
$y$. Because the IMF and Type~II supernova yields are highly
uncertain, $y$ is only constrained to be $0.08\lesssim y\lesssim0.023$
\citep[e.g.,][]{finlator08}; we adopt a mid-range value of $y=0.015$.
We further justify this value and discuss models with a varying yield
in \S\,\ref{sec:igimf}.

The denominator of equation~(\ref{eqn:zgsimple}) includes terms for
metal accretion ($\zetaa$), metal expulsion ($\zetaw$), and dilution
from gas ($\alpha\fg$).  The metallicity-weighted mass-loading factors
$\zetaa$ and $\zetaw$ in equation~(\ref{eqn:mzdot}) describe the
relative rates at which metals are being accreted and expelled from
the system, and are defined as
\begin{eqnarray}
\zetaa &\equiv& \frac{\zigm}{\zg}\times\frac{\macc}{\msfr} = \left(\frac{\zigm}{\zg}\right)\etaa,\;\mbox{and}\label{eqn:zetaa}\\
\zetaw &\equiv& \frac{\zw}{\zg}\times\frac{\mwind}{\msfr} = \left(\frac{\zw}{\zg}\right)\etaw.\label{eqn:zetaw}
\end{eqnarray}
The metallicity of accreting gas, $\zigm$, is typically taken to be
zero, though SPH simulations indicate that due to previous episodes of
enrichment of the intergalactic medium (IGM) from metal-containing
galaxy outflows, the effective $\zigm$ may be non-negligible
\citep{finlator08, oppenheimer09b}.  Because a self-consistent model
of an enriched IGM will be based on the evolution of the \mzr, we will
for now take $\zigm$ and thus $\zetaa=0$, though we will return to the
ramifications of this assumption in \S\,\ref{sec:otherimp}.  The wind
metallicity, $\zw$, is often assumed to be the ISM metallicity
\citep{finlator08,erb08}, giving $\zetaw=\etaw$.  However, $\zg$ is
simply a lower-limit to the possible outflow metallicity (if the wind
is driven by supernovae, then it can be metal-enriched relative to the
ambient ISM, but not metal-depleted).  The actual wind metallicity
will depend on the fraction $\fe$ of the outflow that is entrained
interstellar gas, which has a generic metallicity $\zg$, and the
fraction $1-\fe$ that is from newly exploded supernovae and therefore
has a metallicity $\zmax\sim 0.1$ \citep{woosley95}.  The wind
metallicity is thus
\begin{equation}\label{eqn:zwind}
\zw = (1-\fe)\zmax + \fe\zg,
\end{equation}
where we note that $\fe$ may vary with galaxy mass and must satisfy
$0\leq \fe< 1$.

In this model, as long as galaxies have a given $\fg$-$\mstar$
relation (\S\ref{sec:gas}), how they got that gas (i.e., $\etaa$ and
$\etaw$) is irrelevant.  However, for any given wind model $\etaw$,
the accretion rate as a function of the star formation rate $\etaa$ is
uniquely determined.  We explore this point and its implications in
Appendix~\ref{app:gas}.

By finding combinations of the yield, outflow strength, and gas
fractions that combine as stated on the right-hand side of
Equation~(\ref{eqn:zgsimple}) to give $\zg(\mstar)$ on the left-hand
side, we can explicitly reproduce the observed \mzr.  This is the tack
we take in \S\,\ref{sec:results}.

\subsection{Connecting galaxies and halos}\label{sec:halos}
A number of methods have been developed to empirically connect galaxies
to halos.  One straightforward approach is the cumulative matching of
galaxy ($n_{\rm gal}$) and halo ($n_{\rm halo}$) number counts
\citep{vale04,shankar06,conroy09a}, i.e.,
\begin{equation}
n_{\rm gal}(>\mstar)=n_{\rm halo}(>\mhalo)\, .
\label{eqn:cummatch}
\end{equation}
Assuming that each halo (and subhalo) contains a galaxy,
equation~(\ref{eqn:cummatch}) determines the average mapping between
halo mass and galaxy mass.

\begin{figure}
\includegraphics[width=0.48\textwidth]{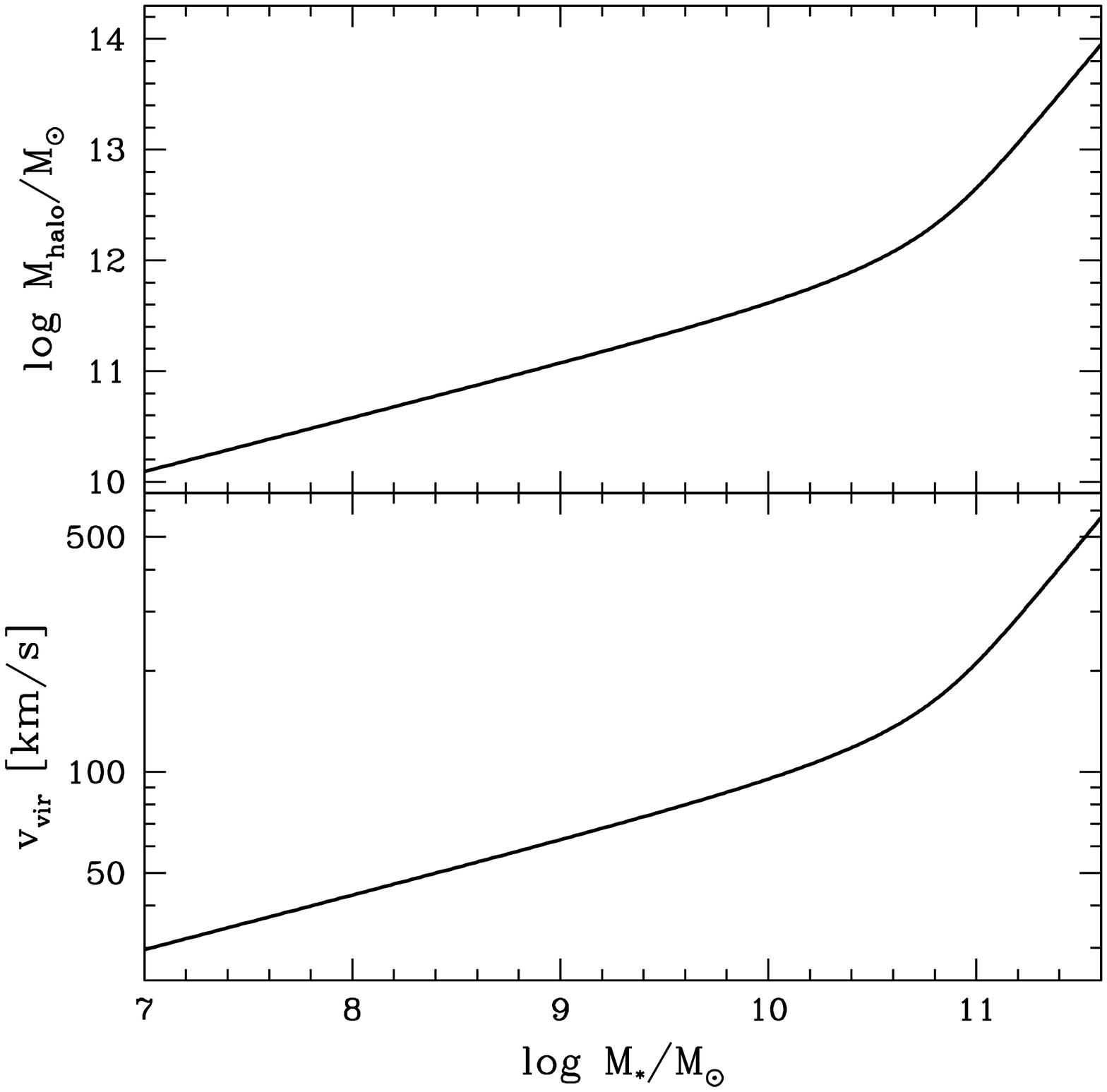}
\caption{\label{fig:halos}Halo mass, $\mhalo$, and virial velocity,
  $\vvir$, as a function of stellar mass, $\mstar$, at $z=0$\
  \citep{moster09}.  See \S\,\ref{sec:halos} for more details.}
\end{figure}

We adopt one of the latest determinations of the $\mstar$-$\mhalo$\
relation by \citet[][top panel of Figure~\ref{fig:halos}]{moster09},
\begin{eqnarray}
\frac{\mstar}{\mhalo} &=& 0.0633(1+z)^{-0.72}\\
&\times& \left[\left(\frac{\mhalo}{M_{{\rm h},0}}\right)^{-1.06-0.17z}+\left(\frac{\mhalo}{M_{{\rm h},0}}\right)^{0.556(1+z)^{-0.26}}\right]^{-1},\hspace*{-1in}\nonumber
\label{eqn:moster}
\end{eqnarray}
with the zero point increased by 0.05 dex to correct from a
\citet{kroupa01} to a \citet{chabrier03a} IMF \citep{bernardi10}, and
where
\begin{equation}
\log M_{{\rm h},0}/\msun = \left[\log 11.88\right](1+z)^{0.019}\, .
\end{equation}
The \citet{moster09} $\mstar$-$\mhalo$\ mapping is in good agreement
with constraints from galaxy-galaxy lensing, galaxy clustering, and
predictions of semi-analytic models.  Following the scaling relations in
\citet[][and references therein]{tonini06}, we have verified that
equation~(\ref{eqn:moster}) yields a Tully-Fisher relation
\citep{tully77} consistent with the more recent calibrations by
\citet{pizagno07}, as long as the dynamical contribution of the dark
matter within a few optical radii is less than the one predicted by a
pure \citet*{navarro96} mass profile, in line with many other studies
\citep[e.g.,][]{salucci07}.  Also note that the subhalo masses quoted by
\citeauthor{moster09}\ refer to {\em unstripped}\ quantities, which
represent more reliable indicators of the intrinsic potential well in
which satellites formed.

The \citeauthor{moster09}\ relation is in broad agreement with other
studies, such as the ones by \citet{guo11} and \citet{shankar06},
although the latter relied on a stellar mass function based on
dynamical mass-to-light ratios that cannot be directly used in the
present study based on SDSS stellar masses.  Despite the different
techniques adopted, most of the studies find consistent results on the
$\mstar$-$\mhalo$\ relation, especially in the mass range of interest
here, i.e., star-forming galaxies with stellar mass $\lesssim 2\times
10^{11}\msun$ and hence halos with mass $\lesssim 5\times 10^{12}\,
h^{-1}\msun$ \citep{firmani09,more10}.  We caution that
\citet{neistein11} have recently described how the $\mstar$-$\mhalo$\
relation could be quite different from expected by the basic abundance
matching technique. Allowing satellite galaxy masses to depend both on
host subhalo mass at infall {\em and} on the friends-of-friends (FOF)
group mass, many distinct galaxy-halo correlations are found to
satisfy all basic statistical and clustering constraints. In
particular, successful models are found where satellite galaxies are
hosted by much lower mass halos than we assume here.

If winds depend on the potential well depth of the halo rather than the
mass itself, then the halo virial velocity $\vvir$ is more relevant than
$\mhalo$.  Roughly speaking,
\begin{equation}
\vvir^2 \sim \Phi \sim \frac{G\mhalo}{R_{\rm halo}},
\end{equation}
where the dependence of the halo radius $R_{\rm halo}$ on the halo mass
is a function of both cosmology and the structure of the halo
\citep{lokas01,ferrarese02,loeb03,baes03}.  We connect $\vvir$\ to
$\mhalo$ via
\begin{eqnarray}
v_{\rm vir}&=&\left(\frac{G\mhalo}{R_{\rm
vir}}\right)^{1/2}\nonumber\\
&=&112.6\left(\frac{\mhalo}{10^{12}\,\msun}\right)^{1/3} \label{eqn:Vvir}\\ 
&\times&\left[\frac{\Omega_m}{0.25}\frac{1}{\Omega_m^z}\frac{\Delta}{18\pi^2}\right]^{1/6}\left(1+z\right)^{1/2}\,
{\rm km\, s^{-1}}, \nonumber
\end{eqnarray}
where the mean density contrast (relative to the critical density)
within the virial radius $R_{\rm vir}$ is $\Delta=18\pi^2+82d-39d^2$,
with $d\equiv \Omega_m^z-1$, and
$\Omega_m^z=\Omega_m(1+z)^3/\left[\Omega_m(1+z)^3+\Omega_{\Lambda}\right]$
\citep{bryan98,barkana01}.  The bottom panel of Figure~\ref{fig:halos}
shows how $\vvir$ varies with stellar mass in this model.  We have
verified that our $\mstar$-$\vvir$\ relation is in good agreement with
the $\mstar$-$v_{200}$\ relation recently derived by \citet{dutton10b}.

\section{Models of the Mass-Metallicity Relation}\label{sec:results}
We now turn to what is required to reproduce the observed
\mzr. Rearranging equation~(\ref{eqn:zgsimple}), we find
\begin{eqnarray}
\frac{y}{\zg} - 1 &=& \zetaw-\zetaa + \alpha \fg,\label{eqn:zetaFg}
\end{eqnarray}
where we hereafter take $\zetaa=0$ (i.e., metal accretion is
negligible; see \S\,\ref{sec:otherimp} for a discussion of this
choice).  Expressed this way, the metallicity $\zg$ is related
explicitly to a sum of $\zetaw$ (a term describing outflows) and
$\fg=\mg/\mstar$ (a term describing the galaxy gas content).
Equation~(\ref{eqn:zetaFg}) is the principal theoretical result of
this paper, connecting gas-phase metallicities to gas fractions,
outflows, and accretion.  Functionally, one can use
equation~(\ref{eqn:zetaFg}) to find working models for a given
$\zg(\mstar)$ in several ways:
\begin{enumerate}
\item\label{it:trial} Assume $y$ and $\zg(\mstar)$ are known; use trial
  and error to find combinations of $\zetaw(\vvir)$ and $[\alpha
  \fg](\mstar)$ that satisfy equation~(\ref{eqn:zetaFg}).
\item\label{it:mg} Assume $y$, $\zg(\mstar)$, and $\zetaw(\vvir)$ are
  known; solve for ${\rm d}\log \mg/{\rm d}\log\mstar$ in
  equation~(\ref{eqn:zgfull}) and integrate to find $\mg(\mstar)$.
\item\label{it:zetaw} Assume $y$, $\zg(\mstar)$, and $\mg(\mstar)$ are
  known; equation~(\ref{eqn:zetaFg}) then says $\zetaw=y/\zg - 1 -
  \alpha \fg$.
\end{enumerate}
Method~\ref{it:trial} works well for developing intuition regarding
tensions in the data and theoretical wind models, while
methods~\ref{it:mg} and \ref{it:zetaw} yield models that exactly
produce the observed \mzr.  In \S\,\ref{sec:constanty}, we explore
models with a constant yield $y=0.015$, focusing in
\S\,\ref{sec:bestfit} on what constraining the model to match observed
gas fractions implies about the efficiency of metal expulsion.  In
Appendix~\ref{app:results}, we go into some of the more subtle details
of how different scalings of $\zetaw$ with $\vvir$ are and are not
consistent with observed gas fractions.  In particular, we show that
the no-winds model ($\zetaw=0$) requires gas fractions that are $\sim
0.3$\,dex higher than observed for all galaxy masses, implying that if
the yield is constant, then the \mzr\ is direct evidence of outflows.
Finally, in \S\,\ref{sec:igimf}, we consider how variable yields
affect our conclusions.

\subsection{Models with constant yield}\label{sec:constanty}
\subsubsection{Choice of $\fg$-$\mstar$ relation}\label{sec:fgmorph}
\begin{figure*}
\includegraphics[width=\textwidth]{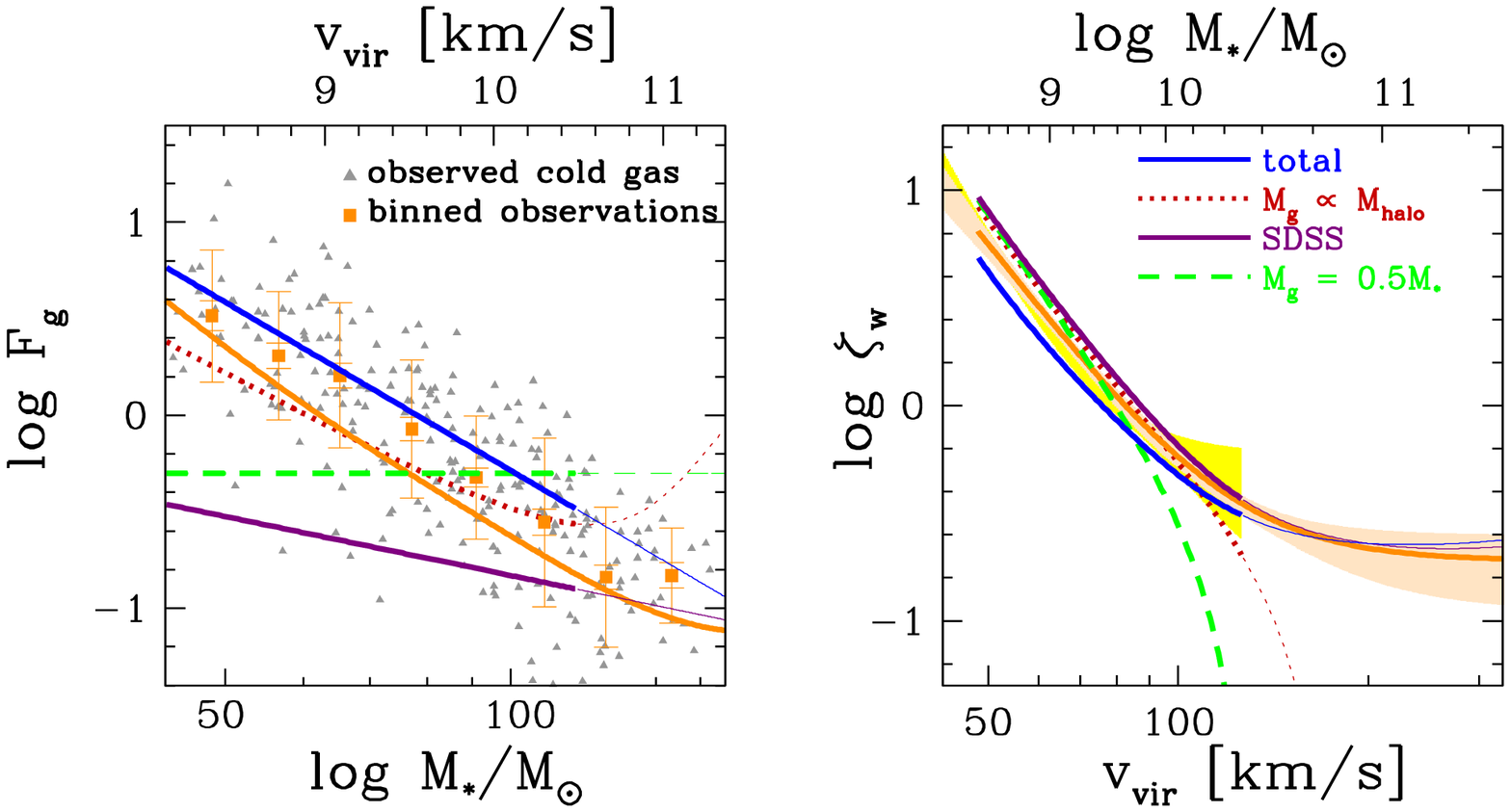}
\caption{\label{fig:mzrT04gas} Required $\zetaw$ to reproduce the T04
  \mzr\ with varying gas fraction relations: total (blue, solid),
  $\mg=0.5\mstar$ (green, dashed), inverting the K-S law from SDSS data
  (purple, solid), and $\mg=0.05(\Omega_b/\Omega_m)\mhalo$ (red,
  dotted); see \S\,\ref{sec:gas} for the motivations behind these
  relations. {\em Left}: Gas fractions as a function of stellar mass.
  The grey triangles in are the gas fractions plotted in
  Figure~\ref{fig:Fg} \citep{mcgaugh05,leroy08,west09,west10} and the
  orange squares are the binned data (\S\,\ref{sec:gas}).  {\em Right}:
  $\log\zetaw$ as a function of virial velocity corresponding to the gas
  fractions in the left panel.  The orange lines are the best-fitting
  models based on the binned data (see Figure~\ref{fig:chisqT04}); the
  beige and yellow shaded regions in the right-hand panel show the
  1-$\sigma$ range in $\zetaw$ for the entire mass range and
  $\log\mstar\leq 10.5$, respectively. }
\end{figure*}

Figure~\ref{fig:mzrT04gas} shows outflow models $\zetaw(\vvir)$ for
different given $\fg$-$\mstar$ relations [method~\ref{it:zetaw}].  As
discussed in \S\,\ref{sec:gas}, we consider the total gas fractions
(blue, solid lines), $\mg=0.05(\Omega_b/\Omega_m)\mhalo$ (red,
dotted), gas fractions as inferred by inverting the Schmidt law for
SDSS galaxies (purple), and $\mg=0.5\mstar$ (green).  The
$\mg\propto\mhalo$ model is included because it might provide a
natural explanation for the observed turnover in the \mzr\ near $M^*$.
We find that for the observed normalization of $\fg(\mstar)$, the {\em
  slope} of the gas fraction relation is largely irrelevant in setting
the \mzr\ morphology.  That is, $z=0$ galaxies have little enough gas
that the \mzr\ is shaped by how $\zetaw$ rather than $\fg$ scales with
galaxy mass.  This can be seen visually in the right-hand panel of
Figure~\ref{fig:mzrT04gas}: at low masses, even the flat gas fraction
relation has approximately the same $\zetaw$ slope as those models
with steep $\fg$-$\mstar$ relations.

\subsubsection{Best-fit models}\label{sec:bestfit}
\begin{figure*}
\includegraphics[angle=270,width=\textwidth]{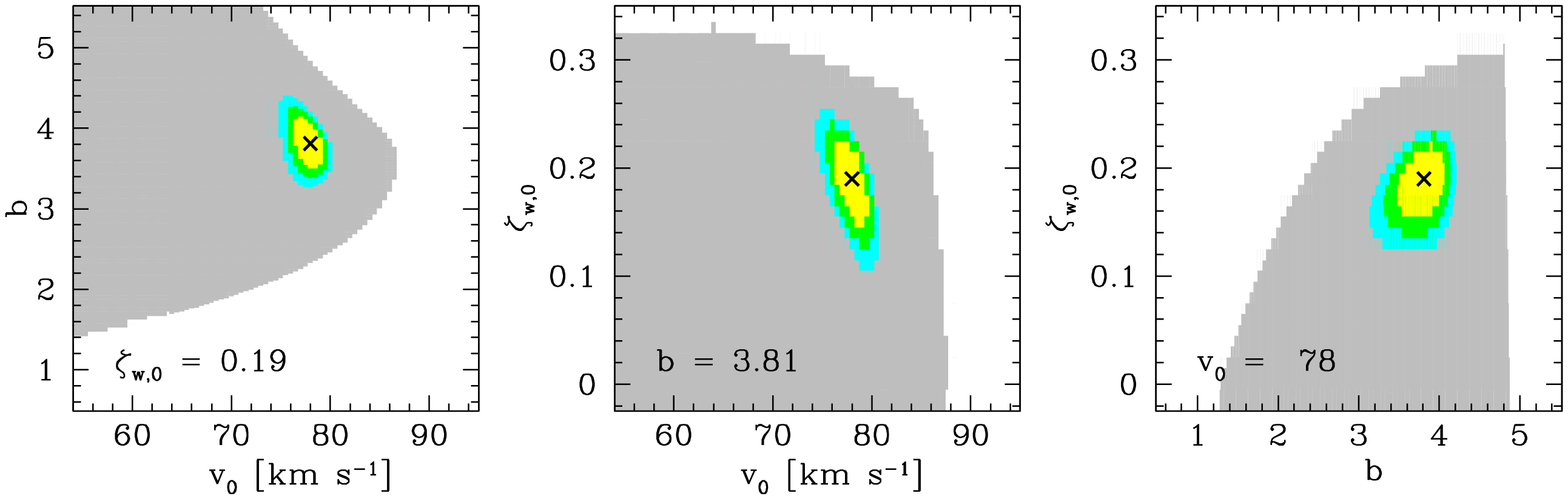}
\caption{\label{fig:chisqT04}$\Delta\chisq$ contours for the T04 \mzr\
  with $\zetaw=(v_0/\vvir)^{-b}+\zeta_{{\rm w},0}$. The black ``X''
  marks the parameters with the lowest $\chisq$; the yellow, green,
  cyan, and grey regions denote solutions with $\Delta\chisq \leq
  1$-$\sigma$, $1\mbox{-}\sigma<\Delta\chisq\leq 2$-$\sigma$,
  $2\mbox{-}\sigma<\Delta\chisq\leq 3$, and $\Delta\chisq>3$-$\sigma$,
  respectively, using the $\Delta\chisq$-to-$\sigma$ conversion from
  \citet{nr}.  The white regions correspond to unphysical ($\mg\leq0$)
  models.}
\end{figure*}

We quantify what $\zetaw(\vvir)$ scalings are required in order to
reproduce the \mzr\ while remaining consistent with the observed gas
fractions by using method~\ref{it:mg}: by taking a given $\zetaw$ we
can compare the corresponding $\fg$ to binned gas fractions
(\S\,\ref{sec:gas}, Table~\ref{tbl:Fgbins}) to calculate a $\chisq$.
Parameterizing $\zetaw$ as $(v_0/\vvir)^{-b}+\zeta_{{\rm w},0}$, the
best-fit model for the T04 \mzr\ is
$\zetaw=(78\kms/\vvir)^{-3.81}+0.19$, as shown in
Figure~\ref{fig:chisqT04}. We show the $\Delta\chisq$ contours for 1-,
2-, and 3-$\sigma$ using the $\Delta\chisq$-to-$\sigma$ conversion
from \citet{nr}\ for 5 degrees-of-freedom (8 data points and 3
parameters).  The best-fit values do not change significantly if the
dispersion about the mean is used instead of the uncertainties when
calculating $\chisq$, and we safely consider the points and errors for
the binned data to be uncorrelated because the measurements for
individual galaxies do not depend on one another.  We bin $\fg$\
instead of $\zg$\ because the \mzr\ has been more rigorously measured
than the $\fg$-$\mstar$\ relation. The white regions in
Figure~\ref{fig:chisqT04} correspond to models that are unphysical
because they require negative gas fractions.  The best-fit models are
always close to the border between physical and unphysical regions in
parameter space, reflecting the fact that gas fractions at $z=0$ are
relatively small; it takes only a small change in $\zetaw$ to go from
a small $\fg$ to a negative one.

\begin{figure*}
\includegraphics[width=\textwidth]{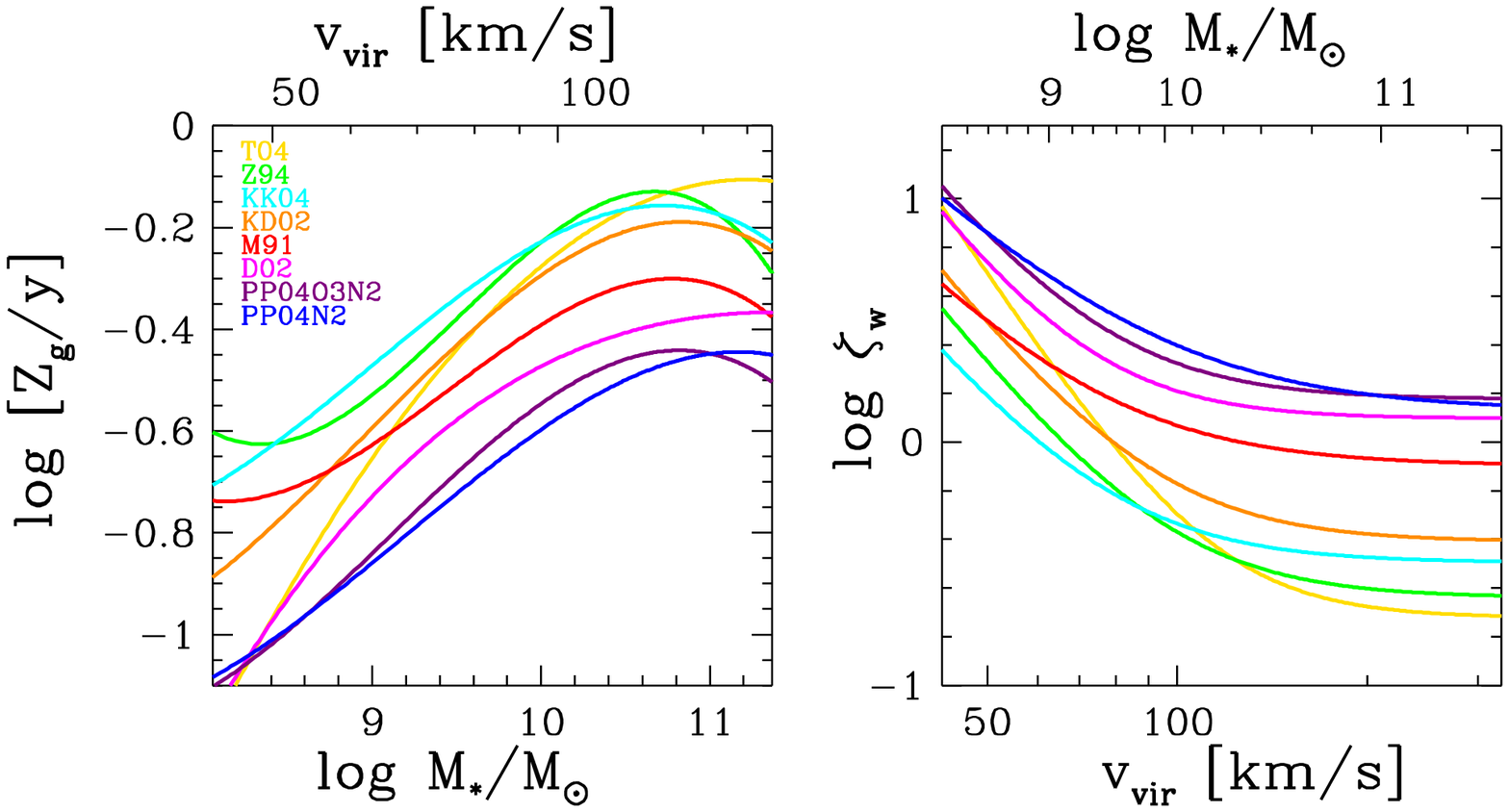}
\caption{\label{fig:bestzetaw} {\em Left:} The mass-metallicity relation
as derived from different metallicity indicators (\S\,\ref{sec:mzrs},
\citealt{kewley08}), relative to the nucleosynthetic yield
$y=0.015$. {\em Right:} The corresponding best fitting
$\zetaw=(v_0/\vvir)^{b}+\zeta_{{\rm w},0}$ under the requirement that
the models' gas fractions are consistent with observations. The $\zetaw$
parameters are listed in Table~\ref{tbl:bestzetaw}.}
\end{figure*}

\begin{table}
\centering
\begin{tabular}{lrrr}\hline\hline
ID & $v_0$ & $b$ & $\zeta_{{\rm w},0}$ \\\hline
T04 & $78.0$ & $3.81$ & $0.19$ \\ 
Z94 & $63.5$ & $3.20$ & $0.23$ \\ 
KK04 & $55.5$ & $3.04$ & $0.32$ \\ 
KD02 & $71.0$ & $3.18$ & $0.39$ \\ 
M91 & $73.0$ & $2.47$ & $0.77$ \\ 
D02 & $79.0$ & $3.42$ & $1.25$ \\ 
PP04O3N2 & $90.0$ & $3.15$ & $1.50$ \\ 
PP04N2 & $111.8$ & $2.31$ & $1.35$ \\\hline
\end{tabular}
\caption{Best-fit parameters for $\zetaw=(v_0/\vvir)^{b}+\zeta_{{\rm
  w},0}$ the fits to the \mzr\ calculated by \citet{kewley08} and listed
  in Table~\ref{tbl:ke08mz} and the binned gas fractions plotted in
  Figure~\ref{fig:Fg}.  These $\zetaw$\ are plotted next to the
  corresponding $\zg(\mstar)$ in Figure~\ref{fig:bestzetaw}.
\label{tbl:bestzetaw}}
\end{table}

The best-fit $\zetaw$\ can be strongly driven by the turnover of the
\mzr\ and change in slope of the $\mstar$-$\vvir$\ relation above
$\log\mstar=10.5$. For example, for the T04 \mzr, if we instead only
consider the data at $\log\mstar<10.5$, the best-fit $\zetaw$ is instead
$(72\kms/\vvir)^{-4.69}+0.41$; that is, the velocity normalization $v_0$
does not change much, but the slope steepens and the constant offset
$\zeta_{{\rm w},0}$ increases.  Whether the best-fit $\zetaw$ shifts to
higher $b$ and $\zeta_{{\rm w},0}$ (T04 and D02), lower $b$ and
$\zeta_{{\rm w},0}$ (M91, Z94, PP04O3N2, and PP04N2), or doesn't change
(KD02 and KK04) when only modeling $\log\mstar< 10.5$ depends on the
subtle details of the particular fit to the \mzr\ under consideration.
In all cases, however, $\Delta\chisq$ for the parameters for the best
fitting $\zetaw$ for a given \mzr\ when the entire mass range is modeled
fall within 1-$\sigma$ of the $\log\mstar< 10.5$ best fitting model for
that indicator (but not necessarily vice-versa, since the best fitting
low-mass model often requires negative gas fractions if extrapolated
above $10^{10.5}\msun$).  The 1-$\sigma$\ range of $\zetaw$ for the T04
\mzr\ is shown by the shaded yellow and beige regions in the right-hand
panel of Figure~\ref{fig:mzrT04gas} for the $\log\mstar< 10.5$ and
entire mass range, respectively.

Other metallicity indicators lead to mass-metallicity relations that
are generally shallower and have a lower normalization than the
\citeauthor{tremonti04}\ \mzr. This translates into $\zetaw+\alpha
\fg$ needing to be larger and to scale slightly less steeply with mass
than seen in Figure~\ref{fig:mzrT04gas}; the best-fit $\zetaw$ for all
of the mass-metallicity relations shown in Figure~\ref{fig:ke08} are
plotted in Figure~\ref{fig:bestzetaw}.  (Detailed example models for
the shallow, low-normalization \citet{denicolo02} \mzr\ are shown in
Figure~\ref{fig:d02}.)  Numerically, observed gas fractions require
$2.3\lesssim b\lesssim 4$; this scaling with $\vvir$ is much steeper
than the canonical models for the unweighted mass-loading parameter
discussed in \S\,\ref{sec:winds}.  Furthermore, $\zetaw$\ must be
large ($\gtrsim 1$) at all relevant masses.  The only way around a
large $\zetaw$ is if a significant fraction of the gas that is
diluting the metals is ionized (and thus not included in the
observations of cold gas our $\fg$-$\mstar$ relations).

In the limit of small $\fg$ and large $\zetaw$, one can see from
equation~(\ref{eqn:zgsimple}) that $\zg\propto\zetaw^{-1}$
\citep{finlator08}.  We are using cubic fits to the \mzr\
(Table~\ref{tbl:ke08mz}, \citealt{kewley08}), but for the relevant mass
range, the \mzr\ has $0.2\lesssim{\rm slope}\lesssim 0.45$ for most of
the relations plotted in Figure~\ref{fig:bestzetaw}.  Our
$\mstar$-$\mhalo$-$\vvir$\ relation (Figure~\ref{fig:halos}) has
$\mstar\propto\vvir^{6}$\ for $\log\mstar\lesssim 10$ (and
$\mstar\propto\vvir^{1.5}$ for $\log\mstar\gtrsim 10.6$).  Thus, the
metallicity $\zg$ is roughly proportional to $\vvir^{1.2}$ to
$\vvir^{2.7}$, implying that for $\zetaw\propto\vvir^{-b}$, $b$ should
be in the range $1.2$ to $2.7$.  The large constant offset $\zeta_{{\rm
w},0}$, however, means that the parameterization presented here (see,
e.g., Figure~\ref{fig:chisqT04}) cannot be directly interpreted in terms
of the simple power-law scalings presented in \S\,\ref{sec:winds}.  We
also caution that $\zetaw\neq\etaw$, and we explore the consequences of
metallicity-weighting the mass-loading parameter below
(\S\,\ref{sec:zwind}).

A crucial step in this analysis is the assignment of virial velocities
to stellar masses.  For example, \citet{finlator08} found that
$\zetaw\propto\vvir^{-1}$ was sufficient to reproduce the $z\sim 2.2$
\mzr\ (which does not differ significantly in slope from the shallow
relations at $z=0$).  In their simulations, however,
$\mstar\propto\mhalo$, which is a shallower relation than our
$\mstar\propto\mhalo^2$, a slope which \citet{moster09} finds to
approximately hold to $z\sim 2$ (see their Figure~14).  Because
$\mhalo\propto\vvir^3$, these differences have extreme consequences
for the interpretation of how $\zetaw$ scales with $\vvir$.

\subsection{Implications of uncertain or varying yields}\label{sec:igimf}
\begin{figure}
\includegraphics[width=0.48\textwidth]{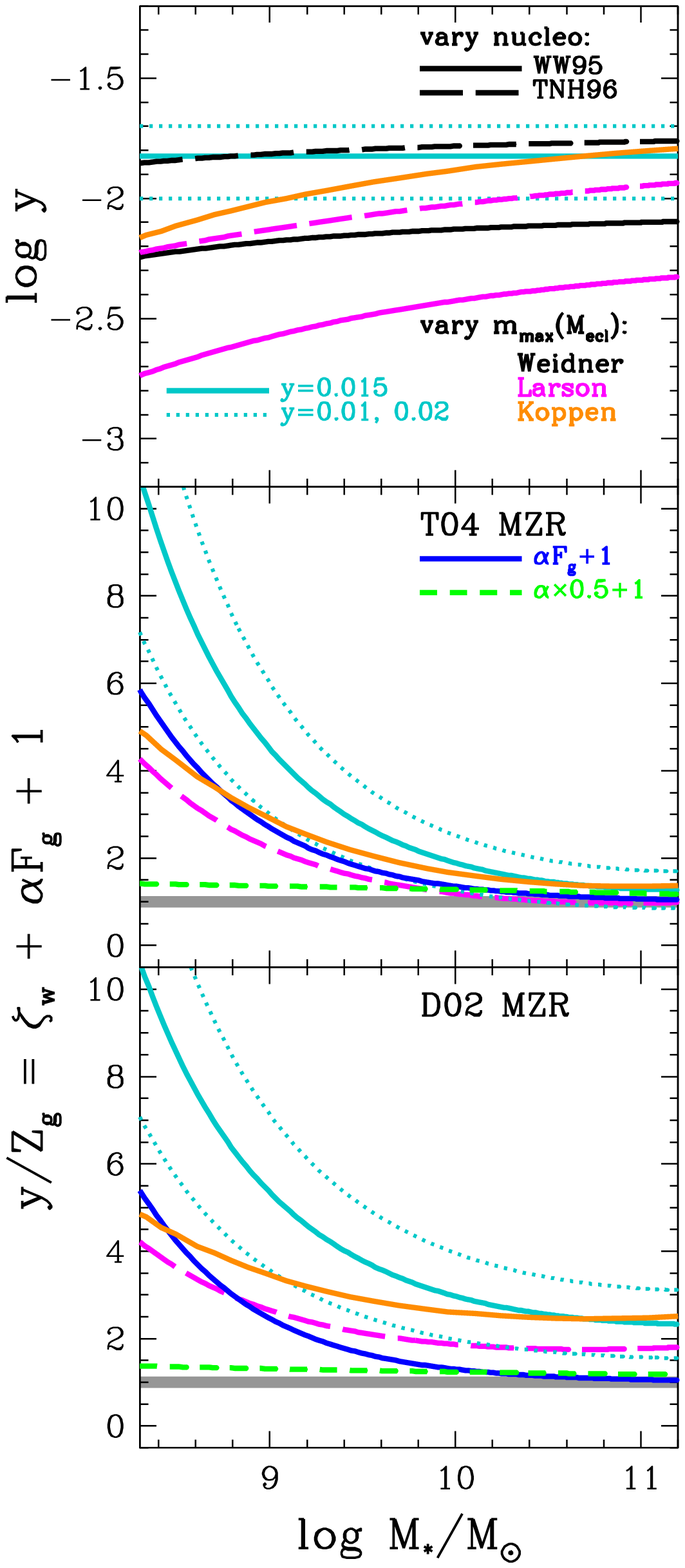}
\caption{\label{fig:igimf} {\em Top}: oxygen yields as a function of
  $\log\mstar$ for different choices of $\mmax(\mecl)$
(\citealt{weidner04}, black; \citealt{larson03}, magenta) and
nucleosynthetic yields (\citealt{woosley95}, solid;
\citealt{thielemann96}, long-dashed).  Effects of different yields on
models of the \mzr\ as $y/\zg=\zetaw+\alpha\fg+1$ for the T04 and D02
mass-metallicity relations are shown in the middle and bottom panels,
respectively. Constant yields are shown in cyan ($y=0.015$, solid;
$\pm 0.005$, dotted); gas fraction line types and colors are the same
as in previous figures, with the total gas fractions as solid blue
lines and $\mg=0.5\mstar$ as short-dashed green lines. Details are given
in \S\,\ref{sec:igimf}.}
\end{figure}

There is increasing evidence that the IMF may vary with star formation
rate, and thus with galaxy mass \citep{meurer09,lee09}.  \citet{kroupa03}
suggest that if all stars form in clusters and if more massive stars
are more likely to form in more massive star clusters, the integrated
galactic initial mass function (IGIMF) depends on the embedded cluster
mass function (the ECMF, $\xi_{\rm ecl}(\mecl)\propto\mecl^{-\beta}$,
where $\mecl$ is the mass of a cluster).  \citet{koppen07} showed that
for certain choices of $\xi_{\rm ecl}(\mecl)$, the $\mmax$-$\mecl$
relation (where $\mmax$ is the most massive star a cluster of mass
$\mecl$ can produce), and SN~II yields, the \mzr\ could be explained
without the need to invoke outflows.  Like many previous models,
however, \citeauthor{koppen07}\ derive stellar and gas masses from
their star formation rates under various assumptions of closed box
with inflows.  We re-examine here the effects of a varying IGIMF on
the \mzr\ using observed gas fractions.  We connect star formation
rates and stellar masses from observations; the median SSFR of
SDSS~DR4 star-forming galaxies can be fit with a power law
\begin{equation}\label{eqn:medssfr}
\log[\msfr/\mstar] = -9.83-0.12(\log[\mstar/\msun]-10),
\end{equation}
as shown as a histogram in Figure~10 of \citet{peeples09a}.

Like \citet{koppen07}, we follow \citet{weidner04} and take the
minimum mass a star cluster can have to be 5\msun\ and the maximum
mass to be governed by the galaxy-wide star formation rate such that
\begin{equation}\label{eqn:meclmax}
\log(\meclmax/\msun) = 4.93 + 0.75\log(\msfr/[\msun\,{\rm yr}^{-1}]).
\end{equation}
We adopt a power-law slope of the ECMF to be $\beta=2$
\citep{lada03, koppen07}.  The IGIMF is thus
\begin{eqnarray}\label{eqn:igimf}
\xi_{\rm IGIMF}(m) \int\limits_{5.0\msun}^{\meclmax}\xi(m\leq\mmax(\mecl))\xi_{\rm ecl}(\mecl){\rm d}\mecl,
\end{eqnarray}
where $\xi(m)$ is the IMF.  The oxygen yield\footnote{The definition
  of yield used in this section, $y\equiv M_{\rm oxy}/\mstar$, is
  slightly different from the one given in Equation~\ref{eqn:yield},
  but that for most purposes these are interchangeable.} as a function
of stellar mass, $y(\mstar)$, will therefore be determined by
$\xi(m)$, the $\mmax(\mecl)$ relation, and the Type~II supernova
yields, as shown in the top panel of Figure~\ref{fig:igimf} (where we
have adopted the \citealt{kroupa01}\ IMF).  The \citet{koppen07}\
yields are shown as the solid orange line.  The purple lines show
models with the \citet{larson03} $\mmax(\mecl)$,
\begin{equation}\label{eqn:larson}
\mmax = 1.2\mecl^{0.45},
\end{equation}
while the black lines show models with the same for models with
$\mmax(\mecl)$ derived from the semi-analytic model of \citet[][c.f.,
the thick solid line in Figure~1 of \citealt{weidner06}]{weidner04};
the \citeauthor{weidner04}\ relation gives a shallower dependence of
the yield on $\mstar$.  The thin solid lines show models derived using
\citet{woosley95}\ nucleosynthetic models ($Z=Z_{\odot}$), while the
dotted lines show the same using the \citet{thielemann96}\ models; as
discussed in detail by \citet{thomas98}, \citeauthor{thielemann96}
gives oxygen abundances that are higher than \citeauthor{woosley95}'s.
Our fiducial value of $y=0.015$ is shown in cyan; this is very similar
to the yields from an IGIMF with the \citeauthor{thielemann96}\ models
with the \citeauthor{weidner04}\ $\mmax$-$\mecl$ relation.

The bottom two panels of Figure~\ref{fig:igimf} show how these
uncertainties in the yield translate to uncertainties in outflows when
modeling the \mzr, where we have plotted $y/\zg$ for the various
yields and for the \citet[][middle panel]{tremonti04}\ and
\citet[][bottom]{denicolo02} mass-metallicity relations.  The thick
grey lines at $y/\zg=1$ denote the boundary below which the yields are
not high enough to produce the observed metallicities.  As shown in
\S\,\ref{sec:model}, the gas fractions and outflows must balance to
give $y/\zg$, we also show $\alpha\fg + 1$ for two gas fraction
models: our fiducial ``total'' gas fractions in blue and a toy
$\mg=0.5\mstar$ model in green.  The difference between the $y/\zg$
curves and the gas fraction curves shows how much outflows are needed.
For example, our fiducial yield of $0.015$ gives $y/\zg$ that is
greater than the $\alpha/\fg+1$ curves at all galaxy masses, with an
decreasing difference at high $\mstar$; these differences are what's
explicitly plotted in Figures~\ref{fig:mzrT04gas}~and~\ref{fig:d02}.
The dotted $y = 0.015\pm 0.05$ lines in Figure~\ref{fig:igimf} show
how these results qualitatively do not change for a large range of
constant yields; this range roughly shows the uncertainty in the yield
from uncertainties in the IMF.  Note that by exploring a range of
normalizations of the \mzr\ in \S\,\ref{sec:constanty}, we are
exploring a range of $y/\zg$---the parameter to which our results are
sensitive.

The closeness of the orange line (the \citeauthor{koppen07} yields)
and the blue line (total gas fractions) in the middle panel shows
that, within reasonable uncertainties, outflows are not strictly
needed in that model.  If the normalization of the \mzr\ is lower,
however, then even with the \citeauthor{koppen07}\ yields, outflows
are required.  This can be explicitly seen by comparing the blue and
orange lines in the bottom panel for the D02 \mzr; moreover, because
the difference between these two curves is greater at larger galaxy
mass, it implies that in this case outflows would have to be {\em
  more} efficent at removing metals from massive galaxies.
Unfortunately, however, the uncertainties in the oxygen production of
core-collapse supernovae, $\mmax(\mecl)$, the IMF, and the
normalization of the \mzr\ all conspire to make drawing strong
conclusions over the nature of outflows in the case of a varying yield
extremely difficult.

\section{Outflow Metallicity and Entrainment}\label{sec:zwind}
Supernova-driven galaxy outflows are comprised of some combination of
supernova ejecta and ambient interstellar medium entrained in the
outflow. The fraction $\fe$ of entrained gas determines wind metallicity
$\zw$.  As mentioned in \S\,\ref{sec:modmzr}, the wind metallicity $\zw$
is usually assumed to be equal to the ISM metallicity $\zg$ when
modeling the \mzr, but if the outflowing supernova ejecta entrains very
little gas (which would dilute the wind metallicity) then $\zw$ could be
much higher than $\zg$.  

We showed in \S\,\ref{sec:results} that models of the observed $z=0$
\mzr\ are more consistent with observations of $z=0$ galaxy gas
fractions when the metallicity-weighted mass-loading factor
$\zetaw\equiv(\zw/\zg)\etaw$ scales steeply with the halo virial
velocity, i.e., $\zetaw=(v_0/\vvir)^{b}+\zeta_{{\rm w},0}$ with
$b\gtrsim 3$.  Theoretical models for how supernovae drive galaxy-scale
outflows, however, generally predict that the {\em unweighted}
mass-loading factor $\etaw\equiv\mwind/\msfr=(\sigma_0/\vvir)^{\beta}$
will scale much more shallowly, with $\beta=1$~or~2
(\S\,\ref{sec:winds}).  Reconciling these disparate scalings therefore
requires that $\zw/\zg$ and hence the wind fluid composition varies with
galaxy mass.

\begin{figure*}
\includegraphics[width=\textwidth]{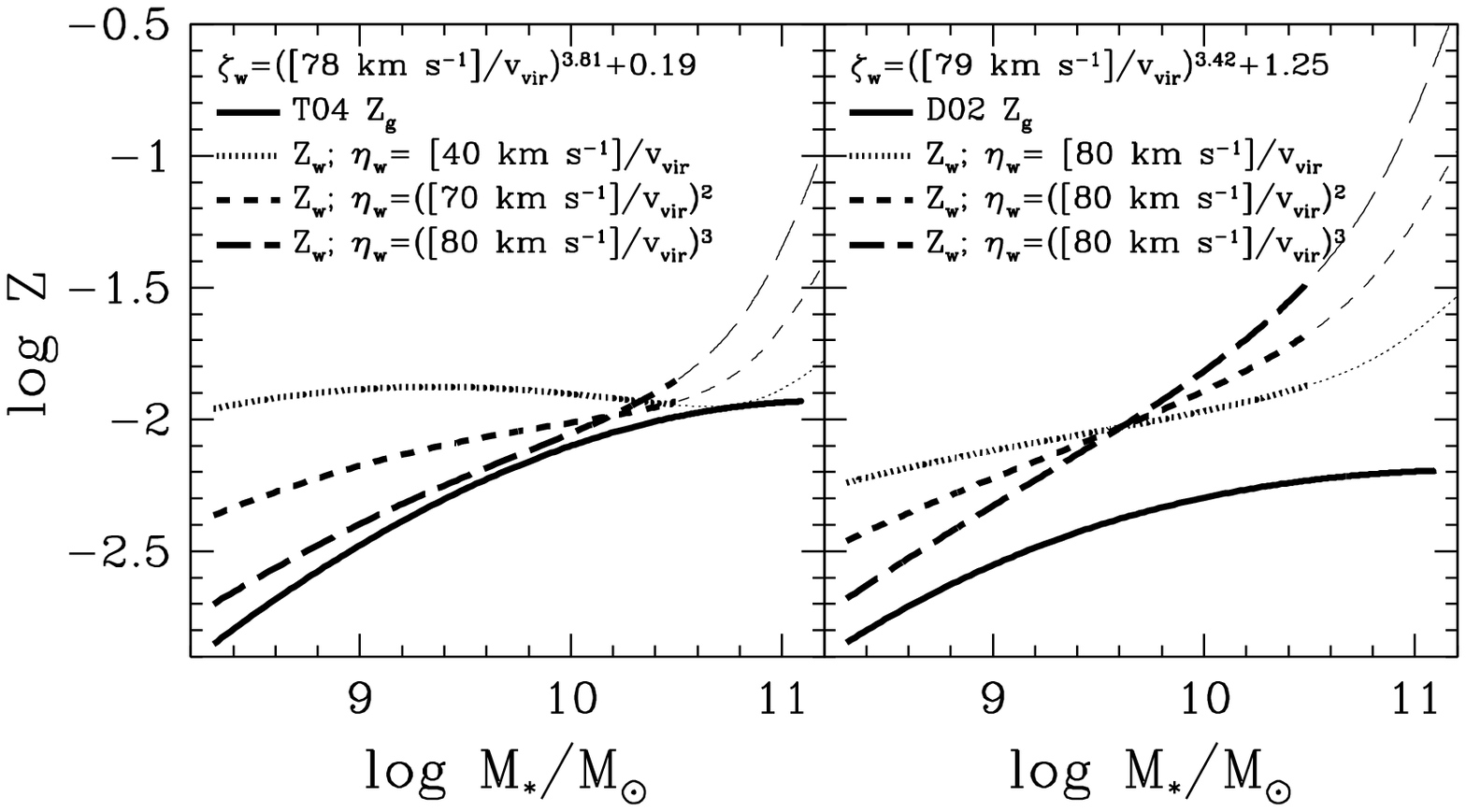}
\caption{\label{fig:zw} Wind metallicities $\zw$ for the best-fit
$\zetaw$ T04 (left) and D02 (right) mass-metallicity relations (see Figure~\ref{fig:bestzetaw}).  The solid line
corresponds to $\zetaw=\etaw$ and therefore $\zw=\zg$ and $\fe=1$;
different scalings for $\etaw=(\sigma_0/\vvir)^{\beta}$ are shown as the
dotted ($\beta=1$), short-dashed ($\beta=2$), and long-dashed
($\beta=3$) lines.  }
\end{figure*} 

For any given $\zetaw$ that reproduces the \mzr, additionally assuming
the form of $\etaw(\vvir)$ uniquely constrains the wind metallicity
$\zw(\mstar)$.  If $\etaw$ is constant with galaxy mass, then
$\zw/\zg$ must increase sharply in lower mass galaxies
\citep{spitoni10}.  Figure~\ref{fig:zw} shows $\zw$ for the best-fit
$\zetaw = (78\kms/\vvir)^{3.81}+0.19$ for the T04 \mzr\ (left) and
$\zetaw = (79\kms/\vvir)^{3.42}+1.25$ for the D02 \mzr\ (right).  The
dotted, short-dashed, and long-dashed lines are for
$\etaw\propto\vvir^{-1}$, $\vvir^{-2}$, and $\vvir^{-3}$, models,
respectively.  If $\etaw$ has a similar scaling with mass as $\zetaw$,
then $\zw\sim\zg$ for all masses.  However, a less steep dependence of
$\etaw$ on $\vvir$ implies that outflow metallicities should depend
less on galaxy mass than $\zg$.  Moreover, determining $\zw$ from
galaxy wind observations has different systematics than determining
$\etaw$, and $\zw$ clearly depends sensitively on the scaling of
$\etaw$.  Figure~\ref{fig:zw} shows how measurements of $\zw(\mstar)$
can therefore be used to place unique constraints on $\etaw$.

Physically, different scalings of $\etaw$ and $\zetaw$ (and thus $\zg$
and $\zw$) indicate that the entrainment fraction $\fe$
(equation~\ref{eqn:zwind}) varies with galaxy mass, offering a clue to
the physics of galaxy outflows.  If, for example, $\fe$ increases with
increasing gas mass (and thus galaxy mass), it would indicate that the
wind fluid does not ``punch'' through a blanketting column density of
gas but instead sweeps up this material and expels it from the galaxy.
On the other hand, $\fe$ decreasing with increasing galaxy mass, would
indicate that the ability of supernova ejecta to collect the
surrounding ISM into the wind fluid depends on the depth of the galaxy
potential well.  We find the former to be the case: to reconcile a
steep $\zetaw$ scaling with a shallower $\etaw$ scaling, then winds
driven from deeper potential wells must be {\em more} efficient at
entraining the ambient ISM than those driven from shallow potential
wells.  We also find that in order to have the normalization of
$\etaw$ be consistent with the normalizations suggested in
\S\,\ref{sec:winds} (i.e., $v_{0}\sim 70$\kms) then the entrainment
fraction must be $\sim 1$, though the exact value is dependent on the
value of $\zmax$.  This is particularly interesting in light of
interpretations of X-ray emitting outflows in which the wind fluid is
almost entirely comprised of supernova ejecta, i.e., $\fe \sim 0$
\citep{strickland09}.  Because iron is primarily not made in Type~II
supernovae and thus likely affected differently by star formation
driven outflows than $\alpha$-elements, stellar [$\alpha$/Fe]
variations could also be used to shed light on the oxygen expulsion
efficiency of galaxy winds \citep{recchi09}.

\section{Summary and Discussion}\label{sec:disc}
\subsection{The approach: modeling a system of galaxies}
We use a simple model of the $z=0$ \mzr\ to place constraints on star
formation driven galaxy outflows. In this formalism
(\S\,\ref{sec:modmzr} and equation~\ref{eqn:zgfull}), the gas phase
(oxygen) metallicity $\zg$ of star forming galaxies is
\begin{equation}
\zg = y\left[\zetaw - \zetaa + \alpha \fg + 1\right]^{-1}, \label{eqn:zgsimple2}
\end{equation}
where $y$ is the nucleosynthetic yield, $\zetaa$ describes accreting
metals, $\zetaw$ describes the efficiency of metal expulsion, $\fg$
describes dilution by gas, and $\alpha$ is a factor of order unity (see
equation~\ref{eqn:alpha}). In the absence of metal accretion
($\zetaa=0$), equation~(\ref{eqn:zgsimple2}) shows that the metallicity
$\zg$ is set by a balance of outflows ($\zetaw$) and gas dilution
($\alpha\fg$), with the normalization set by the nucleosynthetic yield
$y$. This equation represents a general result: each piece can vary with
galaxy mass, halo mass, and redshift.  To the extent that the star
formation history is not bursty, i.e., $\msfr$ varies slowly on
timescales of 10\,Myr then the yield $y$ can be taken as constant with
time, letting equation~(\ref{eqn:zgsimple2}) describe the instantaneous
state of a sequence of galaxies.  Galaxies at $z=0$ are assumed to live
on a hypersurface described by their stellar masses, gas fractions,
metallicities, outflow and host halo properties.  By taking gas
fractions and metallicities from observations, we are therefore able to
uniquely solve for outflow properties in terms of galaxy masses or
metallicities (that are therefore easily comparable to observations) or
in terms of the galaxy potential (and therefore easily comparable to
models of the underlying wind physics).  The only fitting of models to
data in this approach is that of functional forms to observations of the
\mzr\ \citep[e.g.,][]{kewley08} and either models or parameterizations
to gas fractions as a function of stellar mass (\S\,\ref{sec:gas}).
Because there is theoretical uncertainty in which metallicity
indicator(s) to use when calculating the \mzr\ from data, we do not
favor a particular indicator when drawing our conclusions, and
specifically state which constraints come from which pieces of the \mzr.

\subsection{Resulting constraints}\label{sec:constraints}
We consider implications for both the efficiency of star-formation
driven galaxy outflows and for the content of the outflowing material.
The two relevant outflow efficiencies are the efficiency with which a
galaxy expels its metals, $\zetaw\equiv(\zw/\zg)(\mwind/\msfr)$, which
we parameterize as $\zetaw=(v_0/\vvir)^{b}+\zeta_{{\rm w},0}$.  The
second relevant efficiency is that with which a galaxy expels its gas,
the unweighted mass-loading parameter $\etaw\equiv\mwind/\msfr$, which
we similarly parameterize as $\etaw=(\sigma_0/\vvir)^{\beta}$, where
$\beta$ is predicted to be $\sim 1$ or $\sim 2$ with
$\sigma_0=70$--80\kms\ (\S\,\ref{sec:winds}).  The content of the wind
is observed by its metallicity $\zw$, which can be expressed in terms
of the fraction of entrained ISM in the outflow, $\fe$, where
$\zw=(1-\fe)\zmax+\fe\zg$ (equation~\ref{eqn:zwind} in
\S\,\ref{sec:modmzr}).  Under the assumption that $\zigm=0$, we draw
the following conclusions by requiring that viable models reproduce
both the $z=0$ \mzr\ and are consistent with observed cold gas
fractions.

\subsubsection{The necessity of outflows}
Models with no outflows ($\mwind=0\Rightarrow\zetaw=0$) are
inconsistent with observed galaxy gas fractions, if the yield $y$ is
constant.  Specifically, in the absence of winds, the gas masses
needed to dilute the produced metals are higher at all galaxy masses
than the total observed cold gas masses; the magnitude of this offset
is as great as $\sim 0.3$\,dex in $\fg\equiv\mg/\mstar$, depending on
the particular \mzr\ being modeled.

\subsubsection{Constraints from the normalization of the \mzr}
Equation~(\ref{eqn:zgsimple2}) makes it clear that the nucleosynthetic
yield sets the normalization of the \mzr.  From a modeling
perspective, it is useful to consider the \mzr\ normalization relative
to the yield (rather than their absolute values) because the true
nucleosynthetic yield is unknown to a factor of two due to
uncertainties in both the IMF and in Type~II supernova physics
\citep[e.g.,][]{thomas98}.  Likewise, the overall normalization of the
\mzr\ (\S\,\ref{sec:mzrs}) is unknown at the $\sim 0.3$\,dex
level.\footnote{Though neither the nucleosynthetic yield nor the
  normalization of the \mzr\ are well determined, the scatter in
  $\log\zg$ at fixed $\mstar$ is known to be $\pm 0.1$\,dex
  \citep{kewley08}.  In light of the formalism presented here, this
  small scatter implies that either the scatter in both $\alpha\fg$
  and $\zetaw$ are small, or they are highly correlated.}  In the
constant $y$ framework, The normalization of $y/\zg$ sets the value of
the constant offset $\zeta_{{\rm w},0}>0$ (which is set by the
turnover of the \mzr, see below). The typical required velocity
normalization $v_0\sim 70$--$80\kms$ is consistent with expectations.

Low normalization mass-metallicity relations require $\zetaw >1$ for all
relevant masses; if the true nucleosynthetic yield is larger than our
fiducial value ($y>0.015$), then the efficiency with which galaxies
expel metals will have to be even stronger.  Thus if normal quiescently
star forming galaxies are not expelling winds with $\zetaw\gg 1$, then
the data prefer a low nucleosynthetic yield and a high normalization of
the \mzr.  Furthermore, because the \mzr\ shifts to lower normalizations
at higher redshifts, galaxies at these epochs must have either stronger
winds or higher gas fractions than their $z=0$ counterparts.

In \S\,\ref{sec:igimf}, we explored the possibility that the
nucleosynthetic yield $y$ could vary with star formation rate and thus
$\mstar$.  While the possible $y(\mstar)$ relations are still highly
uncertain and the models are much more susceptible to the
$\fg$-$\mstar$ relation, we find that a wide range of such models have
outflows that are more efficient in {\em high} mass galaxies than in
lower mass galaxies.  Observations of such a trend would be compelling
evidence that $y$ varies strongly with galaxy mass.

\subsubsection{Constraints from the morphology of the \mzr}
The morphology of the \mzr\ has two main features: the slope below
$\sim M^*$ and the turnover at higher masses.  In the constant $y$
case, the slope of the \mzr\ largely determines how $\zetaw$\ scales
with galaxy mass, though with some degeneracies with the normalization
and constant offset.  For small $\fg$, as is the case at $z=0$,
$\zetaw$\ should scale roughly as
$\zg^{-1}\sim\mstar^{-0.3}$\,to\,$\mstar^{-0.4}$.  The power-law
scaling of $\zetaw$ with respect to $\vvir$ is typically $b\sim3$.
The required scalings with respect to $\mstar$ are fairly robust,
while the scaling with respect to $\vvir$ depends on our assumed
$\mstar$-$\mhalo$ relation; if this relation is significantly
different from that derived from the abundance matching technique,
then the $\zetaw\propto\vvir^{-3}$ scaling might be able to be
relaxed.  Regardless, this need for a high and mass-dependent wind
efficiency is in broad agreement with previous studies
\citep{dekel03,dutton10a,sawala10,spitoni10}.

The turnover\footnote{The turn-``up'' at low masses for the Z94 \mzr\
  is unphysical and due to the cubic fit to the data.} in the \mzr\ at
$\log\mstar\sim 10.5$ may be an observational artifact of the
metallicity indicators saturating at high $\zg$ (\S\,\ref{sec:mzrs});
however, if oxygen abundances do asymptote to a particular value at
high masses, then this behavior can be used to place strong
constraints on galaxy outflow properties.  Specifically, both the
normalization of the \mzr\ relative to the yield ($\max[\zg/y]$) and
the effects of $\vvir$ increasing sharply above $M^*\sim 10^{11}\msun$
(Figure~\ref{fig:halos}) must be then taken into consideration;
moreover, the interplay between these effects can place stronger
constraints on viable models than just considerations of the \mzr\
below $10^{10.5}\msun$.  Morphologically, a turnover in the \mzr\
means that either $\alpha\fg$ or $\zetaw$ cannot be approximated as a
power-law.  Because cold gas fractions are observed to roughly follow
a power-law with respect to $\mstar$, then $\zetaw$ needs a constant
offset $\zeta_{{\rm w},0}\sim 0.2$--$1.5$, depending on which
indicator is used to calculate the \mzr\ and/or the yield.  In several
cases, if $\mg\propto(\Omega_b/\Omega_m)\mhalo$, then $\zetaw$ can be
described as a power-law; physically, this would imply that galaxies
above $M^*$ have large reservoirs of ionized gas that are able to
efficiently transfer mass with colder, star-forming gas.

If $\zetaw$ and $\etaw$ scale differently, then the
fraction of entrained ISM in the wind fluid will vary with galaxy mass.
Observationally this will be seen as $\zw/\zg$ varying with mass.  As
the morphology of the \mzr\ constrains the scaling of $\zetaw$ with
mass, the scaling of $\zw$ and thus $\etaw$ with mass therefore depends
on the slope of the \mzr.  For example (see Figure~\ref{fig:zw}), for a
fixed $\etaw$, a steep \mzr\ will lead to a shallower $\zw$-$\mstar$
relation than a shallower \mzr\ will. However, since current
uncertainties in the slope of the \mzr\ are smaller than uncertainties
in how (or if) $\etaw$ scales with mass, measurements of $\zw$ across a
large range in galaxy mass, especially above $M^*$, will be particularly
useful for constraining how $\etaw$ (and $\zetaw$) scale.

%%\subsection{Variable yields?}\label{sec:concyield}

\subsection{The role of metal-(re)accretion}\label{sec:otherimp}
At $z=0$, the assumption that accreting material has a negligible
metal content (i.e., that $\zigm=0$ and therefore $\zetaa=0$) may not
be entirely safe.  The IGM is enriched as early as $z>3$
\citep{songaila96,ellison00,schaye03}, and if this material is
re-accreted onto galaxies at later epochs it could have a significant
effect on the shape and normalization of the $z=0$ \mzr.  The
re-accretion of winds (i.e., gas with $\zigm>0$) is a significant
component of accreted gas in cosmological SPH simulations
\citep{oppenheimer09b}.  Though the total accretion rate scales with
halo mass ($\macc\propto\mhalo\propto\vvir^3$, see
Appendix~\ref{app:gas}), the contribution of accreted metals to the
\mzr\ may not scale so steeply \citep{finlator08}.  Moreover, an extra
source of metals $\zetaa$ will imply that the overall efficiency of
outflows (i.e., the amplitude of $\zetaw$) will need to be even higher
than the ones presented here.  However, the reaccretion of wind
material seen in SPH simulations may be sensitive to numerical issues
in the wind implementation; more detailed investigations are needed to
verify the importance of wind-recycling. The metal budget available
for re-accretion depends on both the amount of metals expelled at
higher redshifts and the recyclying timescale.  We will address the
metal content of winds at $z>0$ as implied by the evolution of the
\mzr\ in a later paper.

\section*{Acknowledgements}
We are indebted to David Weinberg and Todd Thompson for numerous
enlightening discussions and comments on the text.  We thank Romeel
Dav\'e, Richard Pogge, Francesco Calura, Francesca Matteucci,
Du\v{s}an Kere\v{s}, Avi Loeb, and Simon White for useful
conversations and Paul Martini, Brett Andrews, and Jonathan Bird for
assistance with obtaining and plotting much of the data presented
here. We are grateful for Andrew West and Barbara Catinella for
providing us with the stellar and gas masses plotted in many of our
figures. We thank the anonymous referee for helpful suggestions in
improving the text. MSP acknowledges support from the Southern
California Center for Galaxy Evolution, a multi-campus research
program funded by the University of California Office of Research. FS
acknowledges support from the Alexander von Humboldt Foundation and
partial support from NASA Grant NNG05GH77G.

%\begin{comment}

%\end{comment}

%%\bibliographystyle{apj} \bibliography{/home/molly/latex/references}

%\clearpage
\appendix

\section{Inverting the K-S Relation}\label{app:ks}
%%%
%
%  K-S RELATION GAS FRACTIONS
%
%%%
The observed Kennicutt-Schmidt \citep[K-S,][]{kennicutt98,schmidt59}
relation is commonly used to indirectly estimate gas masses in
star-forming galaxies in chemical evolution models
\citep[e.g.,][]{spitoni10}, when direct gas masses are expensive (or
currently impossible) to achieve \citep[such as at high
redshifts,][]{erb06b}, or for large samples of galaxies
\citep[e.g.,][]{tremonti04}.  Furthermore, since \tlogoh\ is measured
only in star-forming gas, it is reasonable to consider gas fractions
that trace this same gas.  The purple lines in Figure~\ref{fig:Fg}
(see also Figures~\ref{fig:mzrT04gas}, and \ref{fig:d02}) are the gas
masses we derive from applying the K-S law to star-forming Data
Release 4 SDSS galaxies with $z$-band magnitude errors of $<0.01$\,mag
\citep{brinchmann04,adelman06}.  Specifically, we relate the star
formation rate surface density $\Sigma_{\rm SFR}$ to the gas surface
density $\Sigma_g$ by
\begin{eqnarray}\label{eqn:schmidt}
\Sigma_{\rm SFR}  &\equiv& \frac{\msfr}{A_g} = K_g \Sigma_g^{\alpha}\\
 &=& 1.67\times10^{-4}\left(\frac{\Sigma_g}{1 \msun\,{\rm pc}^{-2}}\right)^{1.4}\msun\,\mbox{yr}^{-1}\,\mbox{kpc}^{-2}\nonumber
\end{eqnarray}
from \citet{kennicutt98}, where we have corrected for the fact that the
\citet{brinchmann04}\ star formation rates are based on a
\citet{kroupa01} IMF while the \citeauthor{kennicutt98}\ relation is
based on a \cite{salpeter55} IMF.  SDSS spectra are taken within a
3\arcsec\ aperture; therefore, to measure {\em total} galaxy properties
(e.g., star formation rates and stellar masses), the fact that the
aperture does not subtend the entire galaxy must be corrected for.  We
therefore consider $\Sigma_{\rm SFR}$ and $\mstar$ both for the full
galaxy-light radius (which we take to be $1.1$ times the 90th percentile
$z$-band isophotal radius $R_{90, z}$) and only within the fiber, i.e.,
we take
\begin{equation}
A_g = \pi R_g^2 = \pi R_{\rm light}^2 =\pi\times\left\{\begin{array}{ll}
  1.1^2\times R_{90, z}^2;&\mbox{solid lines.}\\ R_{\rm
  fiber}^2;&\mbox{dashed line.}\end{array}\right.
\end{equation}
The galaxy gas mass is then simply
\begin{equation}
\mg = \left(\msfr\times \frac{A_g^{\alpha - 1}}{K_g}\right)^{1/\alpha}.
\end{equation}

\section{Outflows, inflows, and star formation: getting the gas masses}\label{app:gas}
\begin{figure*}
\includegraphics[width=\textwidth]{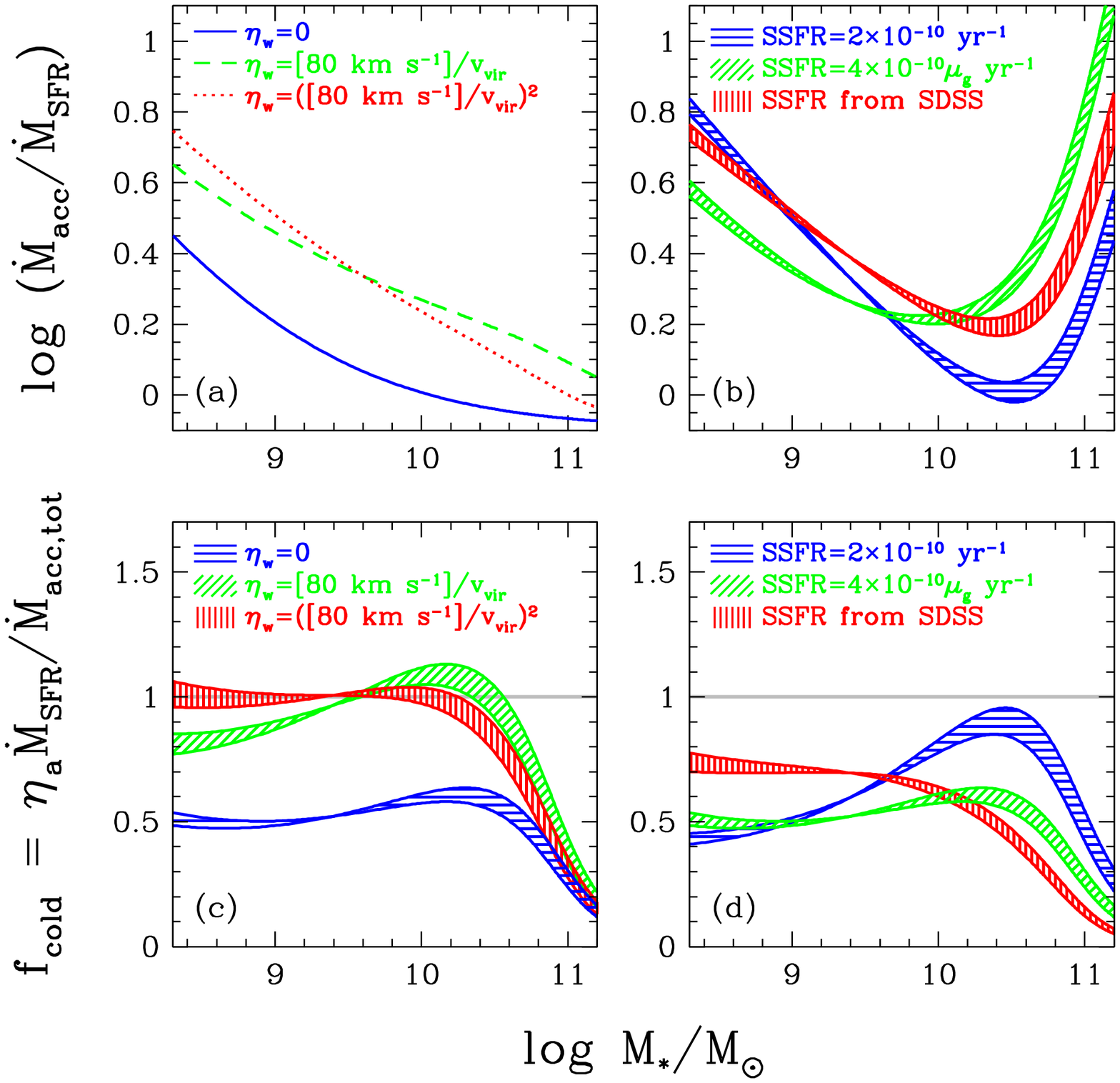}
\caption{\label{fig:fcold} Gas accretion rates, star formation rates,
and cold gas accretion fractions as a function of stellar mass with
varying outflows and specific star formation rates.  All panels assume
the total cold gas fractions described in \S\,\ref{sec:gas}.  Panel (a)
shows how $\etaa$ varies according to equation~(\ref{eqn:etadiff}) for
different $\etaw$ models: no wind (solid blue), a momentum-driven
scaling (green dashed), and an energy-driven scaling (red dotted).  In
all cases, high mass galaxies accrete less gas per unit star formation
than less massive galaxies.  Panel (b) shows the expected range in
$\macctot/\msfr$ between the \citet{neistein06} and \citet{genel08}
$\macc$ models (shaded regions) and with three scalings of
$\msfr/\mstar$ with stellar mass: constant (blue), $\propto\mug$
(green), and the median values from SDSS (red). These $\macctot/\msfr$
are qualitatively similar at low masses to the $\etaa$ shown in panel
(a), but increase rapidly at high masses.  Panels (c) and (d) show the
ratio $\fcold$ of these two estimates, with varying $\etaw$ and the SDSS
SSFRs and with varying the SSFR and no winds, respectively.}
\end{figure*}

As shown in \S\,\ref{sec:modmzr}, 
\begin{eqnarray}
\dot{\mg} &=& \macc - \msfr + \dot{M}_{\rm recy} - \mwind\\
 &=& \msfr(\etaa - 1 + \frecy - \etaw),
\end{eqnarray}
and
\begin{equation}
\frac{{\rm d}\mg}{{\rm d}\mstar} = \frac{\etaa - \etaw - 1 +
  \frecy}{1-\frecy} = \fg (1-\gamma).
\end{equation}
In \S\,\ref{sec:results} we assumed an $\fg$-$\mstar$ relation existed
and that as galaxies evolve they remain on such a relation. Here we
consider, for a given $\etaw$, what implications such a relation has on
the gas accretion rate and how efficiently galaxies are able to turn
this accreted gas into stars.  The above equations imply that the gas
inflow and outflow rates must be balanced by
\begin{equation}\label{eqn:etadiff}
\etaa - \etaw = (1-\frecy)\fg(1-\gamma)-\frecy+1.
\end{equation}
Thus, for a given combination of $\etaw$ and $\fg$, we can uniquely
determine $\etaa\equiv\macc/\msfr$, i.e., the efficiency with which a
galaxy turns its accreted gas into stars.  For example, if the star
formation rate is higher than the accretion rate ($\log\etaa<0$), then
the galaxy is forming stars more quickly than it is accreting gas,
i.e. it is very efficient at forming stars.  We plot $\log\etaa$ for the
no wind, $\etaw=[70\,{\rm km}\,{\rm s}^{-1}]/\vvir$, and
$\etaw=([70\,{\rm km}\,{\rm s}^{-1}]/\vvir)^2$ cases as a function of
stellar mass in the upper-left panel of Figure~\ref{fig:fcold} for the
total gas fraction relation (see Table~\ref{tbl:Fg} and
Figure~\ref{fig:Fg}). Strikingly, $\etaa$\ always decreases
significantly with increasing mass---even in the absence of winds (solid
blue line).  This behavior follows directly from the steepness of the
gas fraction relation (equation~\ref{eqn:etadiff}).  When outflows that
preferentially remove gas from low-mass galaxies
($\etaw\propto\vvir^{-1}$, green dashed line; $\vvir^{-2}$, red dotted
line) are taken into account, $\etaa$ likewise increases and steepens to
compensate.  Therefore, while winds may affect how star-formation
efficiency varies as a function of galaxy mass, they are not necessary
to explain the trend, implying that additional physics is at play.

\begin{figure}
\includegraphics[width=0.48\textwidth]{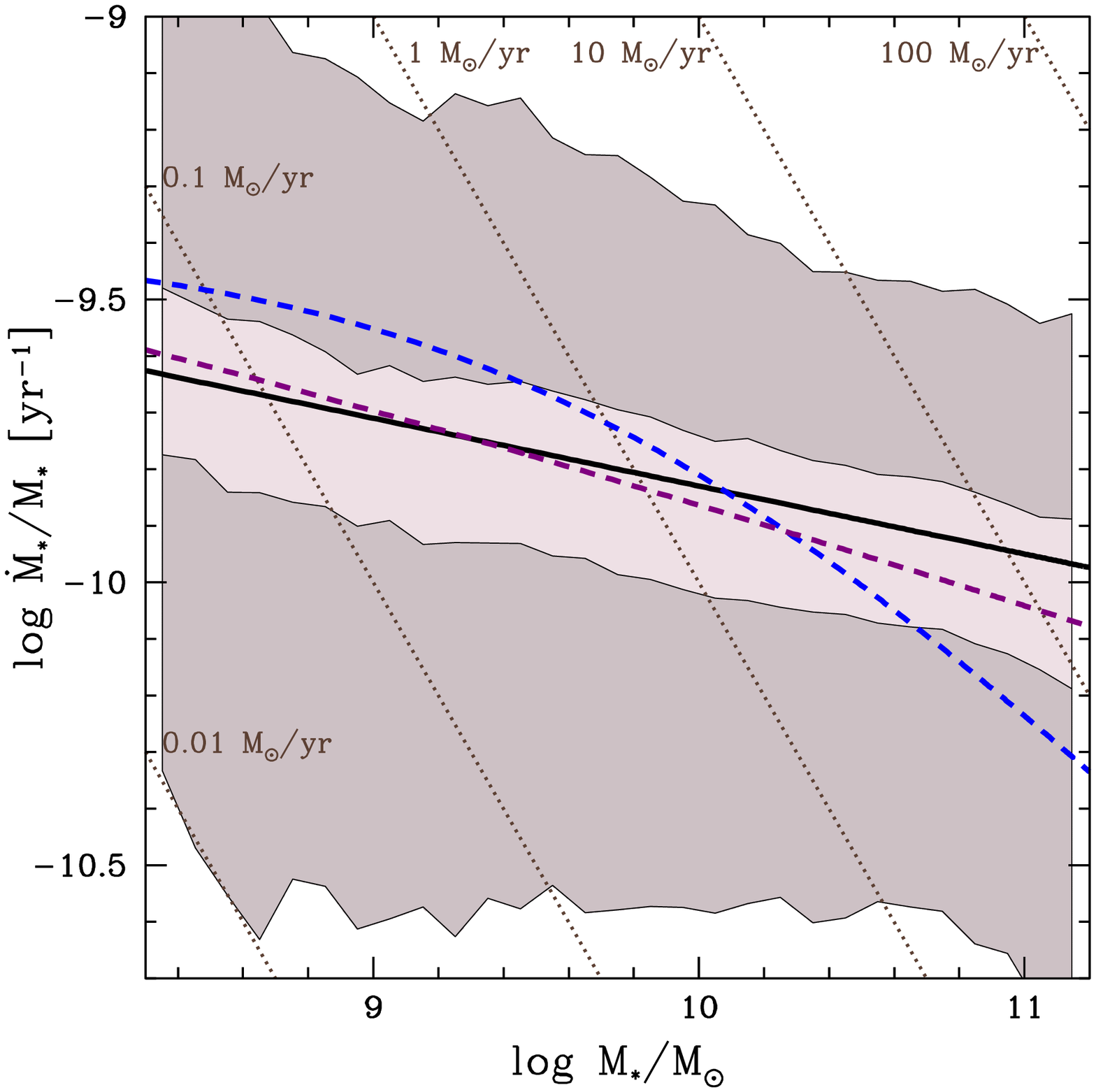}
\caption{\label{fig:ssfr} Specific star formation rates.  The shaded
regions 1- and 2-$\sigma$ dispersions in running bins of $\log\mstar$ of
the aperture-corrected specific star formation rates from SDSS
\citep{brinchmann04}; the black solid line is a power-law fit to median
(equation~\ref{eqn:medssfr}).  The purple dashed is the SDSS $\mug\times
1\times 10^{-9}$\,yr and blue dashed line is the total $\mug\times
4\times 10^{-10}$\,yr; these offsets imply a star formation timescale of
1--2.5\,Gyr.  The shaded regions are dotted lines are constant $\msfr$.
}
\end{figure}

This analysis does not entirely reveal what drives the
$\etaa$-$\mstar$ relation.  However, the nature of $\etaa$ can be
unraveled by appealing to $\msfr$ and $\macc$ from independent
sources.  For example, as shown in Figure~\ref{fig:ssfr}, the median
specific star formation rate (SSFR, $\msfr/\mstar$) in SDSS~DR4
star-forming galaxies decreases with increasing $\mstar$, though there
is large scatter in the SSFR at fixed $\mstar$.  We consider here
three scalings for how the SSFR may vary with $\mstar$. The median
SSFRs from SDSS are shown as the solid line in
Figure~\ref{fig:ssfr}. A physically-motivated way to have the SSFR to
decrease with mass is to postulate that it is proportional to the
total gas fraction, $\mug$.  The blue dashed line shows $\mug\times
4\times 10^{-10}$\,yr for the total gas fractions, while the purple
dashed line shows $\mug\times 1\times 10^{-9}$\,yr for the SDSS gas
fractions (note that the SDSS gas fractions were derived largely from
these same $\msfr$\ and thus this is a somewhat degenerate
comparison).  Finally, we consider a constant SSFR,
$\msfr/\mstar=2\times 10^{-10}$\,yr ($\log[\msfr/\mstar]=-9.7$ in
Figure~\ref{fig:ssfr}).

Using extended Press-Schechter theory, \citet{neistein06} parameterize
the baryonic accretion rate onto halos by
\begin{equation}\label{eqn:macc}
\macctot = 7.23\left(\frac{\mhalo}{10^{12}\msun}\right)^{1.15}\left(\frac{f_b}{0.181}\right)(1+z)^{2.25}\,\msun\,\rm{yr}^{-1},
\end{equation}
where $f_b\equiv\Omega_b/\Omega_m$.  \citet{genel08} find a similar
accretion rate of dark matter onto halos in the Millineum Simulation,
which implies a baryonic accretion rate of
\begin{equation}\label{eqn:genel}
\macctot = 6.34\left(\frac{\mhalo}{10^{12}\msun}\right)^{1.07}\left(\frac{f_b}{0.181}\right)(1+z)^{2.2}\,\msun\,\rm{yr}^{-1}.
\end{equation}
These accretion rates are for matter being accreted into the {\em halo},
not the galaxy, and can be safely considered as upper limits to $\macc$.

The range of $\macctot/\msfr$ allowed between these two $\macc$ models and
three SSFRs (constant, solid; $\propto\mug$, dashed; SDSS median,
dotted) are plotted in the top-right panel of Figure~\ref{fig:fcold}.
At low stellar masses, $\mstar\propto \mhalo^{0.5}$
(equation~\ref{eqn:moster} and Figure~\ref{fig:halos}), which when
combined with the nearly linear mass-dependence of the accretion rate
with halo mass, provides $\macc/\msfr\sim \mhalo/\mstar\sim
\mstar^{-0.5}$, which is the appoximate trend found at $\mstar \lesssim
10^{10}\msun$.  Equations~(\ref{eqn:macc}) and (\ref{eqn:genel}) state
that the overall ``efficiency'' of mass accretion $\macc/\mhalo$ is
roughly constant with halo mass. Therefore, although the host halos of
lower mass galaxies accrete a proportionally equal baryon mass, they are
less capable at converting this gas into stars.  At high masses,
however, the opposite is true: galaxies become more efficient at
converting accreted gas into stars.  For $\mstar \gtrsim 10^{10}\msun$,
$\mstar \propto \mhalo^{0.5}$ (equation~\ref{eqn:moster}), implying
$\macc/\msfr\sim \mhalo/\mstar\sim \mstar$, which is close to the
observed $\etaa$-$\mstar$ slope at high masses.  This combined double
mass-dependent behaviour of $\etaa$ with stellar mass produces the
characteristic ``U'' shape observed in Figure~\ref{fig:fcold}.

The \citeauthor{neistein06}\ and \citeauthor{genel08}\ estimates of
$\macctot$ are for baryonic accretion into the halo.  However, only a
fraction of this infalling gas may be usable for star formation; for
example, if this onfalling gas is shock-heated as it is accreted, then
it will neither be detected in \ionm{H}{I}$+$H$_2$ observations nor
contribute towards star formation (since we are sensitive to $\etaa$
rather than $\macctot$ proper, the gas participating in star formation
is relevant).  Therefore, to better characterize the fraction of gas
that is accreted ``cold''---and therefore able to further cool and form
stars---we combine the estimates of $\macc$, $\msfr$, and $\etaa$,
defining this cold fraction as
\begin{equation}\label{eqn:fcold}
\fcold\equiv\etaa\frac{\msfr}{\macctot},
\end{equation}
where $\macctot$ and $\msfr$ are generally defined. For
illustrative purposes, we let $\macctot$ be defined as in
equations~(\ref{eqn:macc}) and (\ref{eqn:genel}).  Note that to be
physical, $0\leq\fcold\leq 1$.  The lower-left panel of
Figure~\ref{fig:fcold} shows how $\fcold$ varies with different $\etaw$
scalings, assuming the median SDSS SSFRs, while the lower-right panel
shows how $\fcold$ depends on the SSFR in the absence of winds.

\begin{figure}
\includegraphics[width=0.48\textwidth]{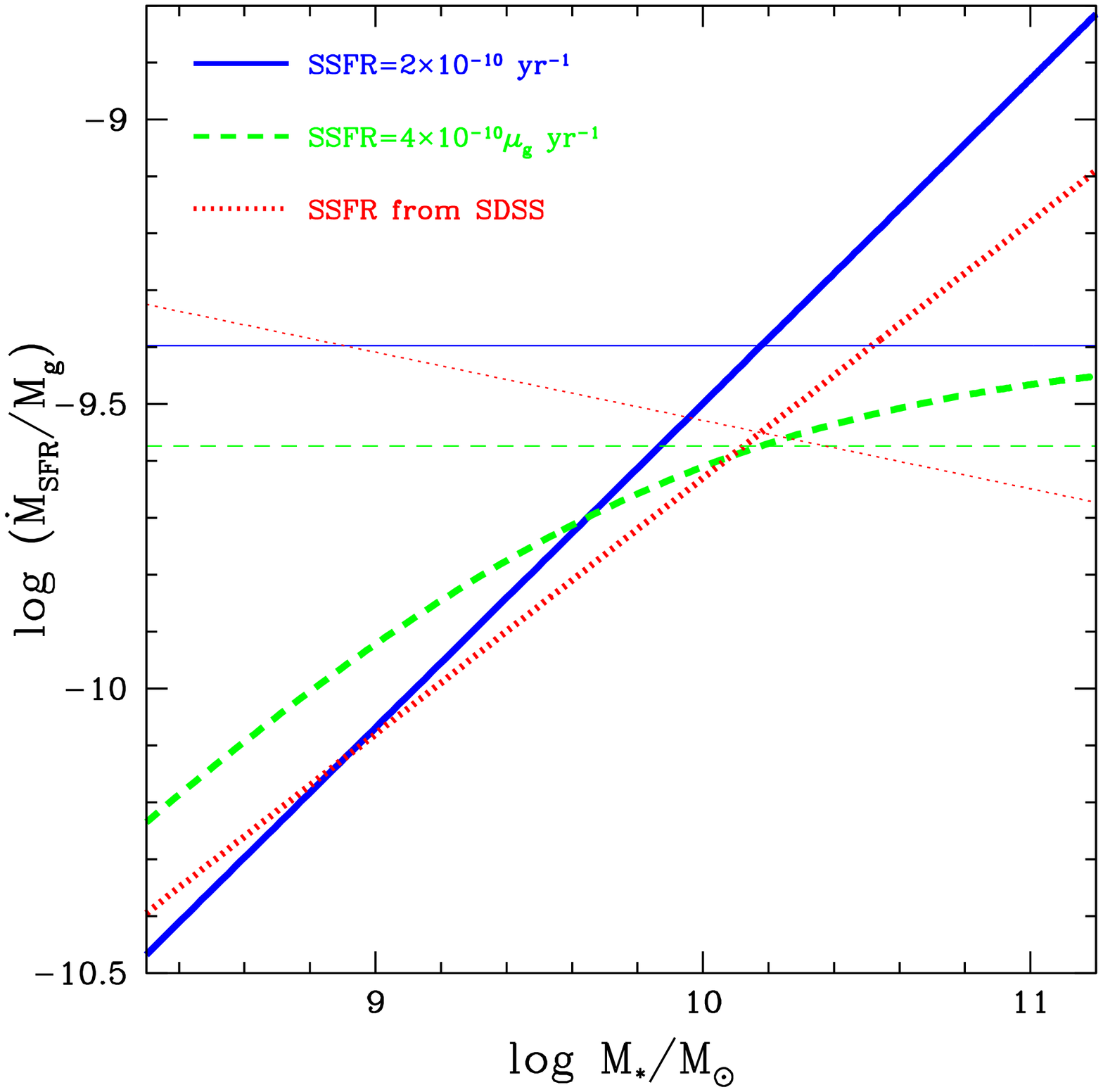}
\caption{\label{fig:sfe} Star formation efficiency $\msfr/\mg$ as a
  function of $\mstar$, taking $\mg$ to be the total cold gas masses
  (thick lines, \S\,\ref{sec:gas}) and $\mg=0.5\mstar$ (thin lines) and
  three choices of the specific star formation rate: constant (solid
  blue lines), $\propto\mug$ (dashed green lines), and the median values
  from SDSS (dotted red lines). In all cases, a steeply decreasing
  $\fg$-$\mstar$ relation is required for the star formation efficiency
  to increase with stellar mass. }
\end{figure}

There are several interesting behaviors in the lower panels of
Figure~\ref{fig:fcold} worth noting.  First, the morphology of
$\fcold(\mstar)$ is fairly robust against variations in the SSFR and
$\etaw$: it is roughly constant, perhaps with a slight rise, for
$\log\mstar\lesssim 10.5$, i.e., below about $M^*$, and then drops
precipitously at higher masses.  Physically, this is a restatement of
galaxies with masses near $M^*$ being more efficient at turning gas
accreted by their halos into stars, relative to either more or less
massive galaxies \citep{shankar06,guo11}.

Second, $\fcold(\mstar)\sim 1$ for low-mass galaxies. At face value,
this would imply that all the accreting gas is available for star
formation.  This closely resembles so-called ``cold-mode'' accretion
scenario in which gas falling into lower mass halos along filaments do
not experience significant shock-heating, thereby easily accreting onto
the central galaxy \citep{dekel06,keres05,keres09a,dutton10a}.  At
higher masses, on the other hand, accreting gas may be shock-heated and
subsequently unable to cool and contribute to star formation.  Despite
this neat picture, however, we find it intriguing that $\fcold(\mstar)$
is so close to unity at low masses. Figure~\ref{fig:Fg} clearly shows
that $\mstar+\mg$ in these same galaxies falls short of accounting for
all of the baryons in the halo by at least a factor of two.  Thus, a
large part of the accreted baryons must be removed from the halo via
strong winds, even if the star formation is reasonably inefficient in
these galaxies, possibly induced by a particularly strong supernova
feedback efficiency.

Finally, Figure~\ref{fig:sfe} builds on this analysis to show the star
formation efficiency, traditionally-defined as $\msfr/\mg$, as a
function of stellar mass for the total cold gas fractions and the three
choices of SSFR.  In all cases, star formation is more efficient in more
massive galaxies: they are forming more stars per unit gas (though see
\citealt{schiminovich10}).  Several previous analyses of the \mzr\ have
suggested that a varying star formation efficiency with galaxy mass is
required in order to reproduce the \mzr\
\citep[e.g.,][]{brooks07,calura09a}.  Figure~\ref{fig:sfe} shows that
this condition is implicitly passed as long as gas fractions are
decreasing with galaxy mass and star formation rates vary reasonably
with stellar mass, as is observed for $z=0$ galaxies.  We note, however,
that with proper choices of $\zetaw$, the \mzr\ is {\em theoretically}
able to be reproduced with a constant $\fg$\ and therefore constant star
formation efficiency.

\section{Deriving the Mass-Metallicity Relation}\label{app:mzr}
The instantaneous change in the stellar mass,
\begin{eqnarray}
\mdotstar &=& \msfr - \dot{M}_{\rm recy}\label{eqn:mstardotexplicit}\\
&=& \msfr(1-\frecy),\label{eqn:dmstardt}
\end{eqnarray}
is given by the creation of stars ($\msfr$) and the rate at which stars
return mass to the ISM when they die, $\dot{M}_{\rm recy}$.  (We include
the mass of stellar remnants in $\mstar$.) The relative rate of these
two effects, $\frecy\equiv\dot{M}_{\rm recy}/\msfr$, depends on the star
formation history and therefore varies somewhat with time; its effect on
our results, however, is small, and we are safe to adopt $\frecy=0.2$.

The gas mass is also regulated by the star formation rate and $\frecy$,
with gas accretion adding gas and outflows removing gas from the
system.  The instantaneous change in  $\mg$ is therefore
\begin{eqnarray}
\dot{\mg} &=& \macc - \msfr + \dot{M}_{\rm recy} - \mwind\label{eqn:mgdotexplicit}\\
 &=& \msfr(\etaa - 1 + \frecy - \etaw),\label{eqn:mgdot}
\end{eqnarray}
where $\macc$ is the gas accretion rate and $\mwind$ is the mass rate of
outflowing gas.  As introduced in \S\ref{sec:winds}, we define the
mass-loading factor $\etaw$ as $\mwind/\msfr$; analogously,
$\etaa\equiv\macc/\msfr$.  The sources and sinks of metals are
essentially the same as for gas, except that each component can have a
different metallicity.  Hence,
\begin{eqnarray}
\dot{M}_{Z} &=& \zigm\macc - \zg\msfr + \zej\dot{M}_{\rm recy} - \zw\mwind\label{eqn:mzdotexplicit}\\
 &=& \msfr(y + \zg(\zetaa - \zetaw - 1)),\label{eqn:mzdot}
\end{eqnarray}
where $\zigm$ is the metallicity of accreting gas, $\zg$ is the ISM
metallicity, $\zej$ is the metallicity of gas being returned to the ISM
by dying stars, and $\zw$ is the metallicity of outflowing gas.  The
yield $y$ is the nucleosynthetic yield, which is defined as the
rate at which metals are being returned to the ISM relative to the
current star formation rate, i.e.,
\begin{equation}\label{eqn:yield}
y = \frac{\dot{M}_{\rm new~metals}}{\dot{M}_{\rm recy}}\times\frac{\dot{M}_{\rm recy}}{\msfr} = \zej\frecy.
\end{equation}
After the first generation of Type~II supernovae explode ($\sim
10^7$\,yr), $y$ is constant for continuous star formation; we adopt a
mid-range value of $y=0.015$.

Since we are interested in the \mzr\ at $z=0$, and not its rate of
change, it is useful to eliminate the time-dependence in
equations~(\ref{eqn:mstardotexplicit}--\ref{eqn:mzdot}).  We assume
galaxies live on a hypersurface of $\mg$, $\mstar$, $\zg$, halo,
accretion and wind properties.  Dividing out the time-dependence in
these equations allows us to solve for the shape of this surface, with
observations setting the amplitude.  Combining
equations~(\ref{eqn:dmstardt}) and (\ref{eqn:mgdot}),
\begin{equation}\label{eqn:dmgdmstar}
\frac{{\rm d}\mg}{{\rm d}\mstar} = \frac{\etaa - \etaw - 1 +
  \frecy}{1-\frecy} = \fg (1-\gamma),
\end{equation}
where we include ${\rm d}\mg/{\rm d}\mstar = \fg (1-\gamma)$ based on
our parameterization of $\fg=\mg/\mstar$ (equation~\ref{eqn:Fg})
introduced in \S\ref{sec:gas}.

The rate of change of the gas phase metallicity $\zg$ is
\begin{equation}\label{eqn:zgdot}
\dot{Z}_g = \frac{{\rm d}\,}{{\rm d}t}\frac{M_Z}{\mg} = \frac{\dot{M}_Z}{\mg} - \frac{\zg}{\mg}\dot{\mg} = \frac{1}{\mg}\Big[\dot{M}_Z-\zg\dot{\mg}\Big].
\end{equation}
We can now combine equations~(\ref{eqn:dmstardt}), (\ref{eqn:mgdot}),
(\ref{eqn:mzdot}), and (\ref{eqn:zgdot}) to find
\begin{eqnarray}
\frac{{\rm d}\zg}{{\rm d}\mstar} &=& 
 \frac{y + \zg\left(\zetaa-\zetaw-1-\frac{\displaystyle \dot{\mg}}{\displaystyle\msfr}\right)}{\mg(1-\frecy)}\label{eqn:Zgexplicit}\\
&&\nonumber\\
&=&  \frac{y + \zg[\zetaa-\zetaw-1-\fg(1-\gamma)]}{\mg(1-\frecy)}.\label{eqn:Zg}
\end{eqnarray}
Equation~(\ref{eqn:Zgexplicit}) can be integrated with
respect to $\mstar$ to find $\zg(\mstar)$.  Furthermore, using the
\citet{kewley08} fits (\S\,\ref{sec:mzrs}, Table~\ref{tbl:ke08mz}), we
can turn the problem around: by assuming we know the \mzr\ (and ${\rm
d}\zg/{\rm d}\mstar$), we can infer the required relation between, e.g.,
$\fg$ and $\zetaw$.  Specifically, by rearranging equation~(\ref{eqn:Zgexplicit}), we find
\begin{eqnarray}
\zg &=& y\Big[\zetaw-\zetaa + \\
    &&\quad\fg(1-\frecy)\left(\frac{{\rm d}\log
      \mg}{{\rm d}\log\mstar} + \frac{{\rm d}\log \zg}{{\rm
	d}\log\mstar}\right) + 1\Big]^{-1}\label{eqn:appzgfull}\nonumber\\
&&\nonumber\\
 &=& y\left[\zetaw - \zetaa + \alpha \fg + 1\right]^{-1}, \label{eqn:appzgsimple}
\end{eqnarray}
where 
\begin{equation}\label{eqn:appalpha}
\alpha\equiv(1-\frecy)\left(\frac{{\rm d}\log
      \mg}{{\rm d}\log\mstar} + \frac{{\rm d}\log \zg}{{\rm
	d}\log\mstar}\right)
\end{equation}
is a factor of order unity, as given in \S\,\ref{sec:modmzr}.

\section{General Models of the Mass-Metallicity Relation}\label{app:results}

\begin{figure*}
\includegraphics[width=\textwidth]{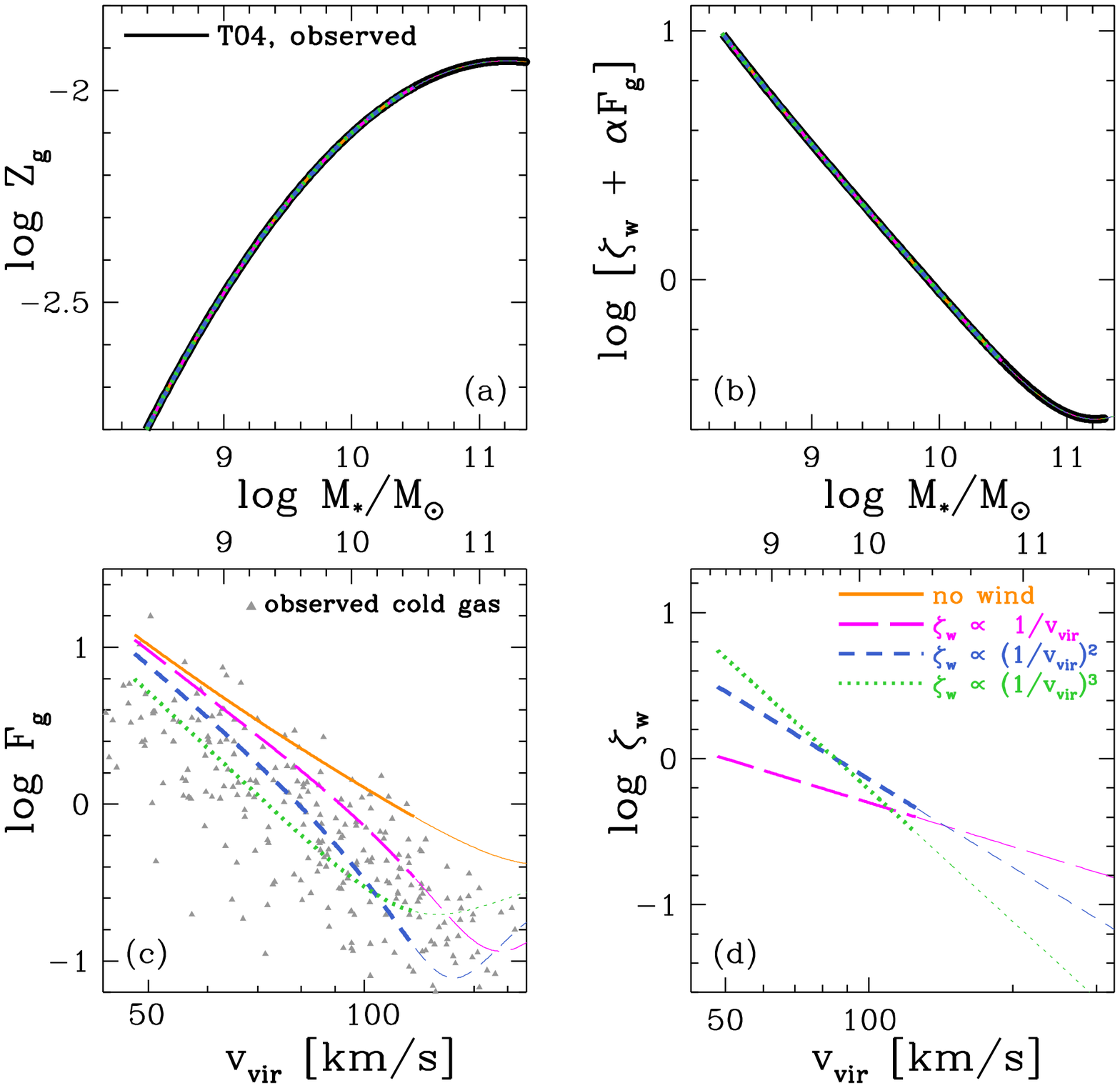}
\caption{\label{fig:mzrT04zetaw} Required gas fractions to reproduce the
  T04 \mzr\ with varying power-law slopes of $\zetaw(\vvir)$: $\zetaw=0$
  (orange, solid), $[50\kms]/\vvir$ (pink, long dashed),
  $([85\kms]/\vvir)^2$ (blue, short dashed), $([85\kms]/\vvir)^3$
  (green, dotted); these normalizations are chosen to give gas fractions
  that are as compatible with the observations as possible.  Note that
  all models fit data in (a) and (b) by construction: panel (a) shows
  the T04 \mzr\ (black, solid) and models (colored lines) while panel
  (b) shows $\log[\zetaw+\alpha\fg]$ as a function of stellar mass for
  the four models (colored lines) and $\log[y/\zg-1]$ for the T04 \mzr\
  in black. Panel (c) shows the model $\log\fg$ as a function of stellar
  mass (colored lines) and the observed gas fractions as grey triangles;
  these are the same observations plotted in Figure~\ref{fig:Fg}
  \citep{mcgaugh05,leroy08,west09,west10}.  The model $\log\zetaw$ as a
  function of virial velocity are plotted in panel (d) (the $\zetaw=0$
  case is unplotted because of the logarithmic $\zetaw$ axis).  }
\end{figure*} 

\begin{figure*}
\includegraphics[height=0.93\textheight]{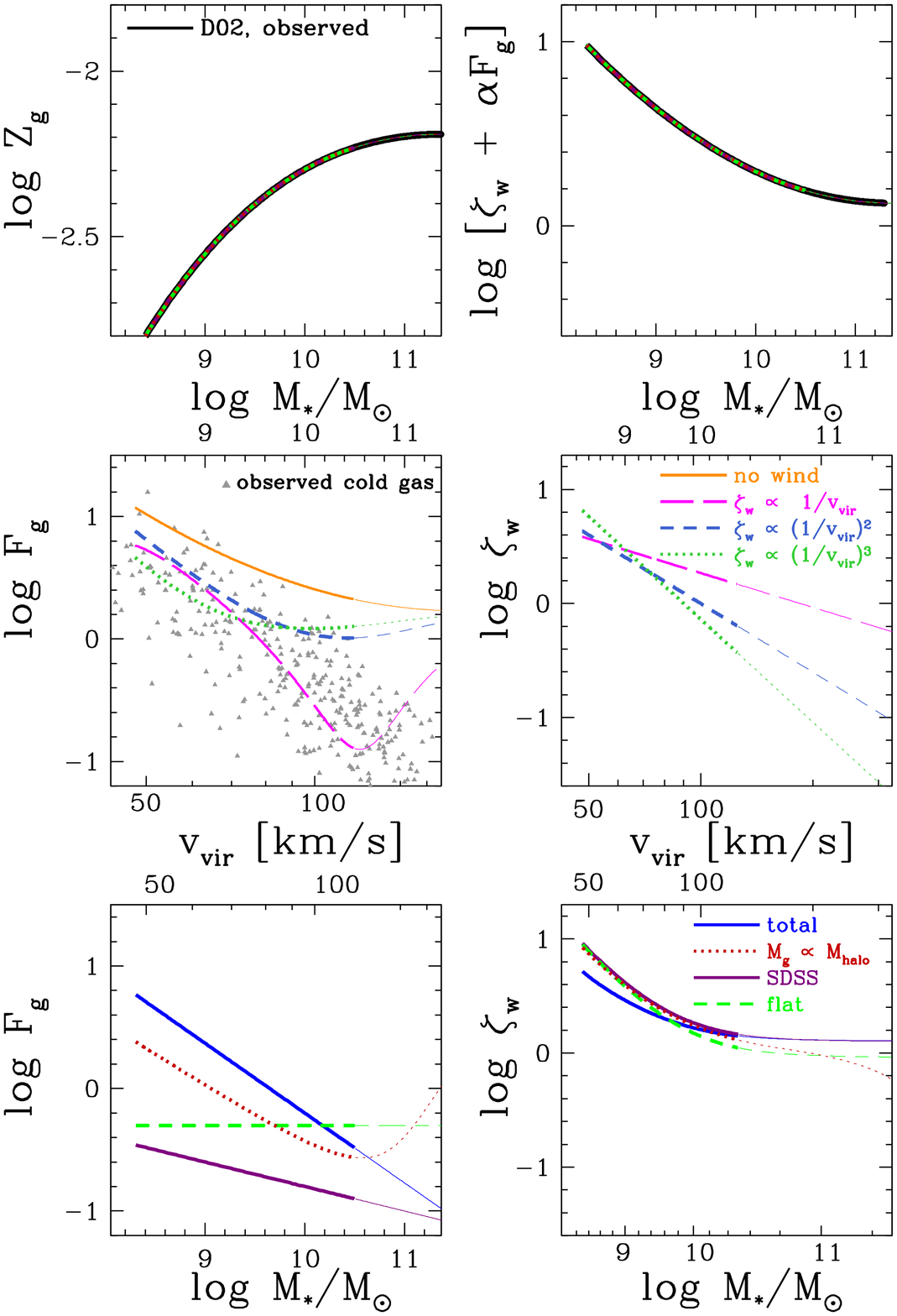}
\caption{\label{fig:d02}Same as Figures~\ref{fig:mzrT04zetaw} and
  \ref{fig:mzrT04gas}, but for the \citet{denicolo02} \mzr. The
  normalizations for $\zetaw\propto\vvir^{-b}$ in the middle two panels
  are chosen to give gas fractions that are as compatible with the
  observations as possible and are: $[185\kms]/\vvir$ (pink, long
  dashed), $([100\kms]/\vvir)^2$ (blue, short dashed),
  $([90\kms]/\vvir)^3$ (green, dotted). }
\end{figure*}

Following methods~\ref{it:mg} and \ref{it:zetaw} in
\S\,\ref{sec:results}, Figures~\ref{fig:mzrT04zetaw} and
\ref{fig:mzrT04gas} show different combinations of gas fractions and
outflow efficencies that explicity repoduce the \citet{tremonti04}\
\mzr; Figure~\ref{fig:d02} shows the same for the \citet{denicolo02}\
relation.  In the top two panels of Figures~\ref{fig:mzrT04zetaw} and
\ref{fig:d02}, the observations are shown as the solid black curves;
the colored lines denote models with different scalings of $\zetaw$
with $\vvir$.  Panel~(a) shows the \mzr\ ($\log\zg$ as a function of
$\log\mstar$).  The models are chosen so that they give $\zetaw+\alpha
\fg$ to equal the observed $y/\zg - 1$ (panel b).  Gas fractions and
$\zetaw(\vvir)$ are plotted in panels~(c) and (d), respectively.
Because of uncertainties in the nucleosynthetic yield, the
normalization of the \mzr, and possible saturation of metallicity
indicators at high \tlogoh\ (\S\,\ref{sec:mzrs}), we will consider
both the \mzr\ across the mass range $8.1\leq\log\mstar\leq 11.3$ and
restricted to below $\mstar\sim 10.5\,\msun$.

The gas fractions needed to dilute the metals in the absence of winds
($\zetaw=0$) are shown as the solid orange line; these gas fractions
are higher by a factor of $\gtrsim 3$ than observed cold gas fractions
in typical $z\sim 0$ galaxies.  For a non-varying yield, outflows are
therefore required in order to keep the observed \mzr\ consistent with
galaxy gas fraction observations.  This conclusion holds even more
strongly for the other mass-metallicity relations plotted in
Figure~\ref{fig:ke08}: in the absence of winds, lower metallicities
imply higher gas fractions.

The other colored lines show the required gas fractions if we assume
$\zetaw\propto\vvir^{-1}$ (pink, long-dashed), $\propto\vvir^{-2}$
(blue, short-dashed), or $\propto\vvir^{-3}$ (green, dotted).  For the
T04 \mzr, both the momentum-driven and energy-driven $\zetaw$ scalings
require $\fg$ to scale more steeply with mass than is observed; lower
normalizations of $\zetaw$ force $\fg$ to asymptote to the no winds
case.  A steeper scaling of $\zetaw$ with $\vvir$, however, leads to
more reasonable gas fractions.

\label{lastpage}

\end{document}